\documentclass{emulateapj}
\usepackage{natbib}
\usepackage[breaklinks,colorlinks,citecolor=blue]{hyperref}

\def\beq{\begin{equation}}
\def\eeq{\end{equation}}

\def\ttco{\mbox{$^{13}$CO}}
\def\twco{\mbox{$^{12}$CO}}
\def\twline{ $^{12}\mathrm{CO} (J = 1 \rightarrow 0)$}
\def\ttline{ $^{13}\mathrm{CO} (J = 1 \rightarrow 0)$}
\def\r25{$r_{\rm 25}$}
\def\rtt{$\mathcal{R}$}
\def\un{\rm \ }
\def\itw{$I_{12}$}
\def\itt{$I_{13}$}
\def\itwc{$I_{12}^{\mathrm{cl}}$}
\def\ittc{$I_{13}^{\mathrm{cl}}$}
\def\Msun{$\mathrm{M_{\odot}}$}
\newcommand{\kms}{$\mathrm{km~s^{-1}}$}
\newcommand{\mc}[1]{\multicolumn{1}{c}{#1}}

\shorttitle{Resolved \ttco \ in Spiral Galaxies}
\shortauthors{Cao et al.}  

\begin{document}

\slugcomment{Accepted for Publication in the Astrophysical Journal} 

\title{CARMA Survey Toward Infrared-bright Nearby Galaxies
  (STING). IV. Spatially Resolved \ttco \ in Spiral Galaxies}
  
\author{
Yixian Cao \altaffilmark{1},
Tony Wong \altaffilmark{1},
Rui Xue\altaffilmark{2,1},
Alberto D. Bolatto \altaffilmark{3},
Leo Blitz \altaffilmark{4},
Stuart N. Vogel \altaffilmark{3},
Adam K. Leroy \altaffilmark{5},
Erik Rosolowsky  \altaffilmark{6}
}

\email{Email for corresponding author: ycao17@illinois.edu}
\altaffiltext{1}{Department of Astronomy, University of Illinois, Urbana, IL 61801, USA}
\altaffiltext{2}{Department of Physics and Astronomy, Purdue
  University, West Lafayette, IN 47907, USA}
\altaffiltext{3}{Department of Astronomy, University of Maryland, College Park, MD 20742, USA}
\altaffiltext{4}{Department of Astronomy, University of California, Berkeley, CA 94720, USA}
\altaffiltext{5}{Department of Astronomy, The Ohio State University, Columbus, OH 43210, USA}
\altaffiltext{6}{Department of Physics, University of Alberta, Edmonton, AB T6G 2E1, Canada}

\begin{abstract}
We present a  $^{13}\mathrm{CO} (J = 1 \rightarrow 0)$ mapping survey of 12 nearby galaxies from the CARMA STING sample.
The line intensity ratio $\mathcal{R} \equiv I[^{12}\mathrm{CO} (J = 1 \rightarrow 0)]/I[^{13}\mathrm{CO} (J = 1 \rightarrow 0)]$ is derived to study the variations in molecular gas properties.
For 11 galaxies where it can be measured with high significance,
the spatially resolved $\mathcal{R}$ on (sub-)kiloparsec scales
varies by up to a factor of 3--5 within a galaxy.
Lower $\mathcal{R}$ values are usually found in regions with weaker $^{12}\rm CO$.
We attribute this apparent trend to a bias against measuring
large $\mathcal{R}$ values when $^{12}\rm CO$ is weak.
Limiting our analysis to the $^{12}\rm CO$ bright regions that are less biased, we do not find $\mathcal{R}$ on (sub)kpc scales correlate with galactocentric distance, velocity dispersion or the star formation rate. 
The lack of correlation between SFR and \rtt \ indicates that the CO optical depth is not sensitive to stellar energy input, or that any such sensitivity is easily masked by other factors.
Extending the analysis to all regions with \twco \ emission by spectral stacking, we find that 5 out of 11 galaxies show higher stacked \rtt \ for galactocentric radii of $\gtrsim 1$ kpc and $\Sigma_{\rm SFR} \lesssim 0.1 \ $\Msun\ yr$^{-1}$ kpc$^{-2}$, which could result from a greater contribution from diffuse gas. 
Moreover, significant galaxy-to-galaxy variations are found in $\mathcal{R}$,
but the global $\mathcal{R}$ does not strongly depend on
dust temperature, inclination, or metallicity of the galaxy.

\end{abstract}
  


\section{Introduction}  \label{sec:intro}
Molecular gas is one of the major components of the interstellar medium,
and its mass is closely correlated to the rate of star formation both on galactic scales \citep[e.g.][]{Kennicutt1998, Gao2004, Evans2006}
and (sub-)kpc scales within galaxies \citep[e.g.][]{Wong2002, Bigiel2008, Rahman2012, Leroy2013}.
The most commonly used tracer of molecular gas in nearby galaxies is the
\twco($J = 1 \rightarrow 0$) transition.
The column density and mass of molecular gas is often estimated from the 
\twco($J = 1 \rightarrow 0$) line intensity (\itw)
by multiplying by a standard \twco-to-$\rm H_2$ conversion factor 
$X_{\mathrm{CO}} \equiv N(\mathrm{H_2})/I_{12} = 2 \times 10^{20}\
 \mathrm{cm^{-2} (K \ km \ s^{-1})^{-1}} $ \citep{Young1982, pbook, Bolatto2013}.
However,  because \twco \  is abundant and
its low level transitions are often optically thick,
\itw \ is only an approximate tracer of column density,
and the $X_{\mathrm{CO}}$ factor can vary depending on 
different underlying physical conditions in the molecular gas. 

\ttco, as a much less abundant ($\sim 1/40$ of \twco \ in the Milky Way) isotopologue, 
is optically thin in most environments,
and can be combined with \twco \ observations to get further information on 
physical conditions in the molecular gas.
Under local thermodynamic equilibrium (LTE), assuming \twco \  and \ttco \ share the same excitation temperature and
originate from the same volume, the opacity of the \ttco \ emission can be derived from the \twco \ and \ttco \ intensities, for optically thick \twco\ that fills the telescope beam.
In Milky Way studies of well resolved giant molecular clouds (GMCs), 
assumptions of LTE and thin to moderately thick \ttco \ are used to derive
the molecular column density and mass \citep[e.g.][]{Goldsmith2008, Roman-Duval2016}.  
More generally, in non-LTE situations, the isotopic line ratio is no longer a simple tracer of opacity and is dependent on the kinetic temperature $T_k$, the density $n({\rm H}_2)$, the \twco\ column density per unit velocity interval $N(\twco)/\Delta v$, and the isotopic abundance ratio [\twco]/[\ttco]; with only one optically thick and one optically thin line it is only feasible to constrain $n({\rm H}_2)$ and $N(\twco)/\Delta v$ by adopting values for $T_k$ and [\twco]/[\ttco] \citep[e.g.,][]{Hirota2010}.

It is often useful to define a \twco \ to \ttco \ flux or intensity ratio \rtt,
$\mathcal{R} \equiv $ \itw $/$\itt,
to characterize  differences between the two types of emission.
Under LTE and optically thin \ttco \ assumptions,

\begin{equation} 
\mathcal{R} \sim \frac{1}{\tau(\mathrm{^{13}CO})} \sim  \frac{[\mathrm{CO}]}{[\mathrm{^{13}CO}]} \frac{1}{\tau(\mathrm{CO})} 
\end{equation}

Explanations for different values of \rtt \  fall into two categories:
changes in the molecular gas opacity due to different underlying physical conditions, 
or variations in  the fractional abundance $[$\twco$]/[$\ttco$]$.
Generally, higher temperature, larger line widths or lower gas column density can reduce
the gas opacity and elevate the \rtt \ value.
The relative isotopic abundance of \twco \ to \ttco \ can be affected by several different
processes. In strong interstellar radiation fields, the less abundant \ttco \ is more
vulnerable to the photodissociation than \twco, so the \twco \ to \ttco \ fractional abundance
will increase and elevate the \rtt \ value.
Another process that may change the isotope abundance is chemical fractionation 
towards \ttco \ at $ \lesssim 35 \rm \  K$ \citep{Watson1976}.
Furthermore, chemical evolution of $^{12}$C and $^{13}$C through galactic nucleosynthesis
can also play an important role in determining \rtt \ \citep{Henkel1993, Prantzos1996}.

Early studies of \ttco \ in nearby galaxies
used single dishes to derive the flux ratio of  \twco \ to \ttco \ of a galaxy,
and found that \rtt \ values are very high ($>20$) in (U)LIRGs which host 
strong starbursts \citep[e.g.][]{Aalto1991,Casoli1992},
while \rtt \ in normal galaxies typically ranges from 5 to 15
\citep[e.g.][]{Y&S1986, S&I1991, Aalto1995, Paglione2001}.
 \rtt \ shows weak correlations with the infrared(IR) color $F_{60}/F_{100}$
\citep{Aalto1991, Crocker2012} and the star formation rate surface density \citep{Davis2014}.
One explanation of these weak correlations
is that  the \twco \  opacity is reduced when feedback from active star formation
heats the interstellar medium (ISM) and increases the velocity dispersion of gas
\citep{Crocker2012,Davis2014}.
However, no strong correlation between \rtt \ and
galaxy properties has been found,
with studies that have
surveyed a considerable number of normal galaxies revealing
a large scatter in \rtt \ \citep{Aalto1995, Vila-Vilaro2015}.

There are also a handful of nearby galaxies that have been imaged in \ttco \
using interferometers \citep[e.g.][]{Turner1992, M64, Meier2004, Aalto2010, Konig2016}.
These mapping studies show spatial variations of  \rtt \ on (sub-)kpc scales within a galaxy;  
the typical value of \rtt \ is $\sim 6$ in
galaxy disks, similar to that in Milky Way disk clouds \citep{Polk1988, Roman-Duval2016}.  
Abnormally high \rtt \ and steep \rtt \ gradients are often found in extreme environments 
such as the nuclei of starbursts \citep{Aalto2010},
or in places where dynamic evolution of gas structures may take place \citep{Huttemeister2000, Meier2004}.
However, spatially resolved \ttco \ studies usually focus on a single galaxy;
whether and how resolved \rtt \ changes in  response to local environmental conditions remains unclear.
A survey of spatially resolved \ttco \ combining multiple galaxies can be used to 
extend the range of local environments  investigated.
Furthermore, galaxy-to-galaxy differences can also contribute to 
systematic differences in \rtt. Therefore, such a resolved survey can be 
used to study the effects of both local environments and global galaxy  properties on \rtt.
   
In this paper, we study  \rtt \ by comparing the \twco \ and \ttco \ emission
over a diverse sample of galaxies from the CARMA STING survey.
We present the observational data in Section \ref{sec:obs}. 
We describe our measurement of the \ttco \ intensity and \rtt  \ in
Section \ref{sec:measure}, as well as the possible bias 
resulting from limited sensitivity.  
In Section \ref{sec:results}, we investigate the dependence of spatially resolved \rtt \ on 
local line width, galactocentric distance, and star formation rate in
each galaxy, and the correlations between global \rtt \ and galaxy properties in the sample. 
We discuss the implications and limitations of this work in Section \ref{sec:discussion}. 

\section{Observations} \label{sec:obs}
\begin{deluxetable*}{l l l r r r c r  c c c}
\tablecaption{\label{table:ginfo} Basic properties of the sample galaxies}
\tablewidth{0pt}
\tabletypesize{\scriptsize}
\tablecolumns{11}
\tablehead{
\colhead{Galaxy}  &  \colhead{R.A.\tablenotemark{a}} & \colhead{Dec.\tablenotemark{a}} &  \colhead{Distance\tablenotemark{b}} &
\colhead{$i$}  &  \colhead{P.A.} & \colhead{$i$, P.A.\tablenotemark{c}} & \colhead{$R_{25}$ \tablenotemark{d}}
&  \colhead{$\log(\rm O/H)$} & \colhead{Metal.\tablenotemark{e}} & \colhead{Morph. \tablenotemark{a}}\\
 & \colhead{(J2000)} &\colhead{(J2000)} & \colhead{(Mpc)} & \colhead{($\arcdeg$)} & \colhead{($\arcdeg$)}
& \colhead{Ref.} & \colhead{($\arcsec$)} &\colhead{$+12$} & \colhead{Ref.} & \colhead{class} }
\startdata 
NGC0772  &   $ 01^{\rm h}59^{\rm m}19^{\rm s}.6 $   &
 $ +19\arcdeg00\arcmin27\arcsec.1 $   &   $ 30.2 $   &   $ 37 $   & 315 & 1
  &   $ 217.3 $   &   $ 8.87 \pm 0.13 $  & 1  &  Sb \\
NGC1569  &   $ 04^{\rm h}30^{\rm m}49^{\rm s}.1 $   &
 $ +64\arcdeg50\arcmin52\arcsec.6 $   &   $ 2.5 $   &   $ 63 $   &  112  & 2
  &   $ 108.9 $   &   $ 8.13 \pm 0.12 $   & 2  & IB \\
NGC1637  &   $ 04^{\rm h}41^{\rm m}28^{\rm s}.2 $   &
 $ -02\arcdeg51\arcmin28\arcsec.7 $   &   $ 9.8 $   &   $ 39 $   &  213 &1
  &   $ 119.5 $   &   $ 8.80 \pm 0.34 $   & 3  &  Sc \\
NGC3147  &   $ 10^{\rm h}16^{\rm m}53^{\rm s}.7 $   &
 $ +73\arcdeg24\arcmin02\arcsec.7 $   &   $ 40.9 $   &   $ 32 $   & 147 & 3 
  &   $ 116.7 $   &   $ 9.02 \pm 0.36 $   & 4 &  Sbc \\
NGC3198  &   $ 10^{\rm h}19^{\rm m}55^{\rm s}.0 $   &
 $ +45\arcdeg32\arcmin58\arcsec.6 $   &   $ 14.0 $   &   $ 72 $   &  215 & 4 
  &   $ 255.3 $   &   $ 8.62 \pm 0.28 $   & 5 &  Sc \\
NGC3593  &   $ 11^{\rm h}14^{\rm m}37^{\rm s}.0 $   &
 $ +12\arcdeg49\arcmin03\arcsec.6 $   &   $ 5.5 $   &   $ 67 $   &   90  & 1 
  &   $ 157.4 $   &   $ 8.29 \pm 0.26 $   & 4  & S0-a \\
NGC4254  &   $ 12^{\rm h}18^{\rm m}49^{\rm s}.6 $   &
 $ +14\arcdeg24\arcmin59\arcsec.4 $   &   $ 15.6 $   &   $ 31 $   &    69  & 5 
  &   $ 161.1 $   &   $ 8.79 \pm 0.34 $   & 5 & Sc \\
NGC4273  &   $ 12^{\rm h}19^{\rm m}56^{\rm s}.1 $   &
 $ +05\arcdeg20\arcmin36\arcsec.0 $   &   $ 36.6 $   &   $ 61 $   &    189  & 1 
  &   $ 70.3 $   &   $ 9.14 \pm 0.20$   & 6 & Sc \\
NGC4536  &   $ 12^{\rm h}34^{\rm m}27^{\rm s}.1 $   &
 $ +02\arcdeg11\arcmin17\arcsec.3 $   &   $ 14.7 $   &   $ 68 $    &    301  & 6
  &   $ 227.6 $   &   $ 8.61 \pm 0.40 $  & 5 &  SABb \\
NGC4654  &   $ 12^{\rm h}43^{\rm m}56^{\rm s}.6 $   &
 $ +13\arcdeg07\arcmin36\arcsec.0 $   &   $ 16.1 $   &   $ 62 $   &   125  & 6 
  &   $ 146.9 $   &   $ 8.83 \pm 0.27 $   & 7 &  SABcd \\
NGC5713  &   $ 14^{\rm h}40^{\rm m}11^{\rm s}.5 $   &
 $ -00\arcdeg17\arcmin20\arcsec.3 $   &   $ 21.4 $   &   $ 33 $   &   203  & 7 
  &   $ 82.7 $   &   $ 8.64 \pm 0.40 $   & 4 & SABb \\
NGC6951  &   $ 20^{\rm h}37^{\rm m}14^{\rm s}.1 $   &
 $ +66\arcdeg06\arcmin20\arcsec.3 $   &   $ 23.3 $   &   $ 46 $   &   138  & 8 
  &   $ 116.7 $   &   $ 8.99 \pm 0.36 $  & 4 &  SABb  \\
\enddata 
\tablenotetext{a}{Data from NED.}
\tablenotetext{b}{Weighted average from NED of redshift-independent distances.}
\tablenotetext{c}{ References for $i$ and P.A. 
(1) Axis ratio and position angle are from $K_s$ (LGA/2MASS
isophotoal) on NED Diameters page, in homogenized units.
Inclination is derived from the axis ratio using Hubble's
(1926) formula with intrinsic flattening of 0.11.   
(2) \citet{2005AJ....130..524M}; (3) \citet{2008MNRAS.388..500E}; 
(4) \citet{2008AJ....136.2648D}; (5) \citet{2008MNRAS.385..553D};
(6) \citet{2006MNRAS.366..812C}; (7) \citet{2006MNRAS.367..469D};
(8) \citet{2009ApJ...692.1623H}.}

\tablenotetext{d}{Semi-major axis from RC3 in arcsec, from NED Diameters page,
  homogenized units.}
\tablenotetext{e}{References for metallicity. 
(1) \citet{2010MNRAS.407.2660A}; 
(2) \citet{2008ApJ...678..804E}; 
(3) \citet{1998AJ....116.2805V};
(4) Calculated from absolute $B$ magnitude using $L-Z$ relations by \citet{Moustakas2010};
(5) Characteristic values from Table 9 in \citet{Moustakas2010}; 
(6) \citet{2008ApJ...673..999P}; 
(7) \citet{2004AA...425..849P}.
}
\end{deluxetable*}

We make use of observations of \twco \ and \ttco \ from the CARMA STING survey, 
which encompasses 23 northern ($\delta > -20\arcdeg$), moderately inclined ($i < 75\arcdeg$) galaxies 
within 45 Mpc selected from the IRAS Revised Bright Galaxy Survey \citep{Sanders2003} 
to sample a wide range of stellar mass.  Previous results from the STING \twco\ data 
have been reported by \citet{Rahman2011,Rahman2012} and \citet{Wong2013}. 
Observations were conducted from 2008 to 2010 using the C, D, and E configurations of CARMA, 
utilizing both 6m and 10m diameter antennas, and yielding a typical
synthesized beam of $\sim$3\farcs5 
(corresponding to scales of $\sim 100 - 700 \rm \ pc $).  
Simultaneous \twco\ and \ttco\ observations were only possible for datasets beginning in mid-2008, 
due to early limitations in the IF bandwidth.  
Furthermore, in several galaxies (NGC 337, 2976, 3486, 3949, 5371) the \ttco\ line was not confidently detected. 
Thus we focus in this paper on the 12 galaxies listed in Table \ref{table:ginfo}. 
We note that equal integration time was spent on \twco\ and \ttco, so the signal-to-noise ratio (SNR) for 
the \ttco\ maps is generally poorer because of the relative weakness of the line.

For most galaxies, observations were conducted using a Nyquist-sampled 19-pointing mosaic pattern 
that provided an effective field of view of 100\arcsec\ in diameter.
For some of the more inclined galaxies (NGC 2976, 3949) the mosaic pattern was compressed along 
the minor axis to improve sensitivity.  
The CARMA receivers were tuned to the CO ($J = 1 \rightarrow 0$) line at 115.2712 GHz 
in the upper sideband and the $^{13}$CO ($J = 1 \rightarrow 0$) line at 110.201 GHz in the lower sideband. 
The spectral resolution was 0.9766 MHz (2.6 \kms), which after Hanning smoothing was effectively $\sim$5 \kms.
However, we typically imaged the data with 10 \kms\ channels to improve the SNR. 
Most galaxies were observed with 3 overlapping 63 MHz windows,
providing a velocity coverage of $\sim 420$ \kms. 
Galaxies with wider \twco \  lines (NGC 772, 3147) were observed in multiple correlator settings across different days,
and therefore have sensitivity and angular resolution varying as a
function of frequency. 
For these galaxies we image each velocity segment separately and
convolve to a common angular resolution 
after deconvolution, while tracking the sensitivity variation 
across the cube for purposes of signal detection.

\begin{deluxetable*} {l c c c c c l r r}[t!]
\tablecaption{\label{table:mom0} Mapping properties of the sample galaxies}
\tabletypesize{\small}

\tablehead{
Galaxy
& \multicolumn{1}{c}{$\theta_{\mathrm{maj}} \times
  \theta_{\mathrm{min}}$ \tablenotemark{a}}
& 
& \multicolumn{3}{c}{Mean noise levels}
&
&\multicolumn{2}{c}{Number of half-beams} 
\\
 &  \multicolumn{1}{c}{$(\arcsec \times \arcsec)$} &
& \multicolumn{3}{c}{$(\rm  K \un km \un s^{-1})$}  &   & 
\multicolumn{2}{c}{ detected with SNR $> 3$ in}  \\
\cline{4-6}   \cline{8-9}  
 & & & \mc{$\left<\sigma_{13}\right>_{\rm Det}$ \tablenotemark{b}}
 & \mc{$\left<\sigma_{13}\right>$ \tablenotemark{c}} 
 & \mc{$\left<\sigma_{12}\right>$ \tablenotemark{d}} 
& &  \colhead{\ttco \ \tablenotemark{e}}  
 & \colhead{\twco \  \tablenotemark{f}} 
} 

 \startdata
 NGC0772  &   $ 5.00 \times 5.00 $   &    &  1.31  &  1.25  &  2.21  &
&  19  &           537  \\
NGC1569  &   $ 4.71 \times 4.31 $   &    &  $ \ldots$ &  0.84  &  1.22  &
&  
           0  &            12 \\  
NGC1637  &   $ 3.48 \times 3.17 $   &    &  1.61  &  1.16  &  1.73  &
&  
          18  &           345 \\ 
NGC3147  &   $ 7.50 \times 7.50 $   &    &  0.66  &  0.56  &  0.80  &
&  
         102  &           420 \\ 
NGC3198  &   $ 3.18 \times 2.82 $   &    &  2.50  &  1.67  &  2.48  &
&  
          15  &           160 \\ 
NGC3593  &   $ 3.87 \times 3.68 $   &    &  2.85  &  2.06  &  3.07  &
&  
          52  &           312 \\ 
NGC4254  &   $ 3.34 \times 2.74 $   &    &  2.05  &  1.85  &  2.35  &
&  
         665  &          2244 \\ 
NGC4273  &   $ 2.87 \times 2.73 $   &    &  3.49  &  2.16  &  3.46  &
&  
          13  &           298 \\ 
NGC4536  &   $ 3.86 \times 3.46 $   &    &  5.35  &  3.57  &  4.01  &
&  
          32  &           158 \\ 
NGC4654  &   $ 3.10 \times 2.90 $   &    &  1.33  &  1.05  &  1.68  &
&  
          84  &          1121 \\ 
NGC5713  &   $ 5.07 \times 4.64 $   &    &  1.24  &  0.87  &  1.25  &
&  
          42  &           348 \\ 
NGC6951  &   $ 9.96 \times 8.53 $   &    &  1.72  &  0.92  &  0.82  &
&  
          12  &           224 \\
 \enddata

\tablenotetext{a}{FWHM of the major and minor axes of the common Gaussian beam
    obtained after convolution of both the \twco \ and \ttco \ cubes.}  

\tablenotetext{b}{Mean $1\sigma$ noise of \itt \ for the \ttco-detections in the M12 mask.} 

\tablenotetext{c}{Mean $1\sigma$ noise of \itt \ for the entire M12 mask.}
 
\tablenotetext{d}{Mean $1\sigma$ noise of \itw \ for the entire M12 mask.}

\tablenotetext{e}{Number of half-beams detected in \ttco \ with SNR $>3$, i.e. 
the \ttco-detections defined in Section \ref{sec:mom0}.}
  
\tablenotetext{f}{Number of half-beams detected in \twco \ with SNR $>3$,  
i.e. both of the \ttco-detections and the \twco-only-detections defined in in Section \ref{sec:mom0}.} 
\end{deluxetable*} 

Data reduction was performed using the MIRIAD package \citep{Sault1995}, 
which applies linelength corrections, passband and phase calibration,
and flux density bootstrapping using a planet or bright quasar.  
Imaging and deconvolution were performed using the MIRIAD tasks INVERT and MOSSDI with 
Briggs' ``robust'' parameter set to 0.5.  
The \twco\ and \ttco\ cubes were then convolved and re-sampled to a
common spatial grid and angular resolution before undertaking further analysis.
A pixel size of $1 \arcsec$ was used, and the common angular resolution is the larger
of the \twco \ and \ttco \ cubes' resolution. 
The FWHM major and minor axes of the common resolution Gaussian beam 
are listed for each galaxy in Table \ref{table:mom0}.


\section{Measurements} \label{sec:measure}

\subsection{Integrated intensity} \label{sec:mom0}

\begin{figure*}[h!]
\hspace{0.7cm}
\epsscale{1.04}
\plotone{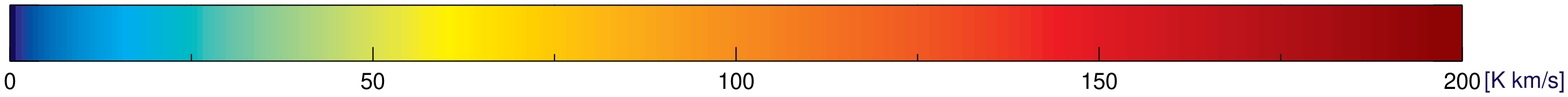}\\
\epsscale{1}
\plotone{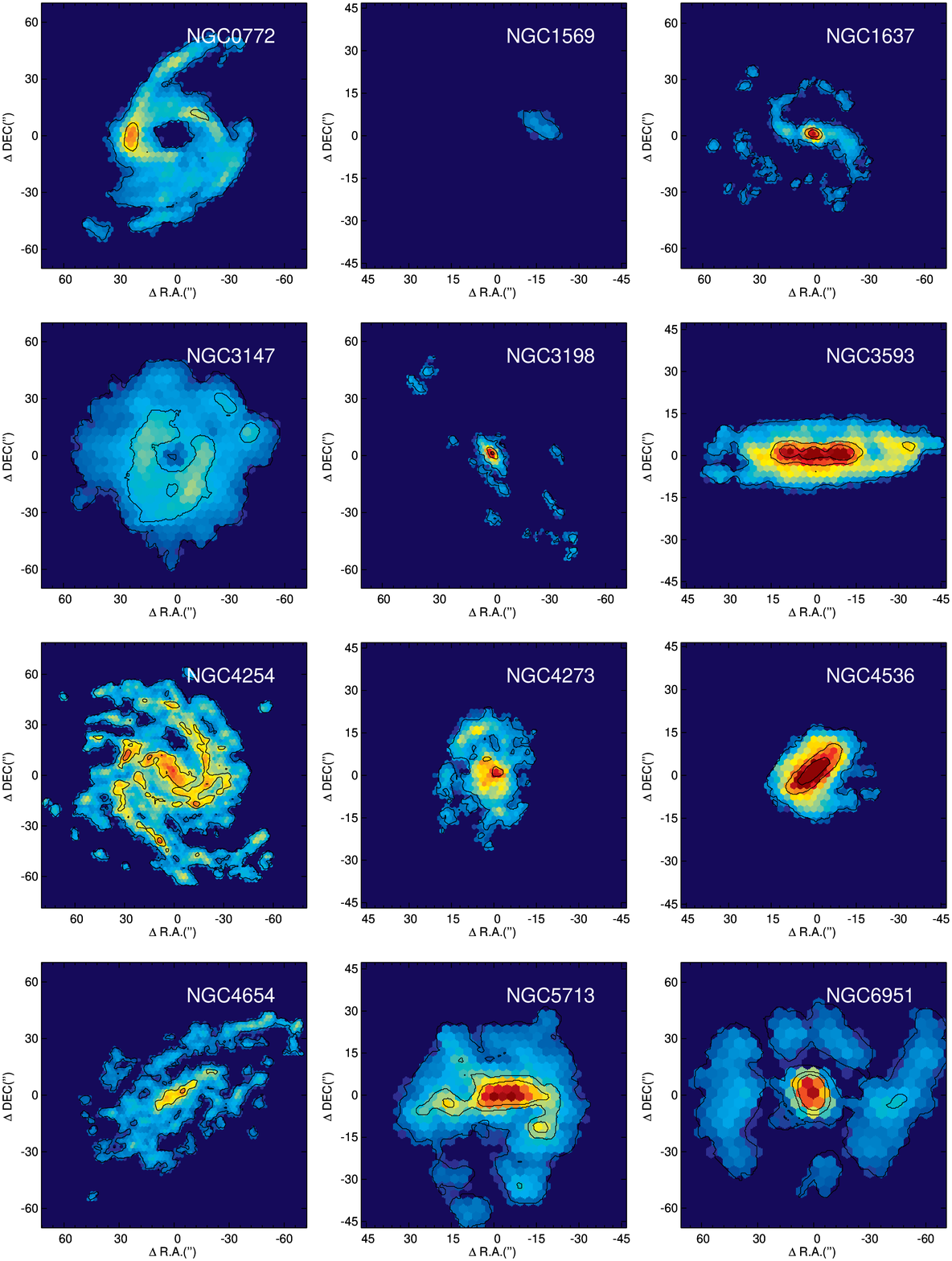}
\caption{\twline \ integrated intensity maps of the 12 STING 
galaxies with \ttline \ detected. 
The maps are resampled to hexagonal grids as described in Section 
\ref{sec:mom0}.  
The contours show SNR of \twline  \ at the levels of $[3, 21, 39]$.} 
\label{fig:12all}
\end{figure*}

\begin{figure*}
\hspace{0.7cm}
\epsscale{1.04}
\plotone{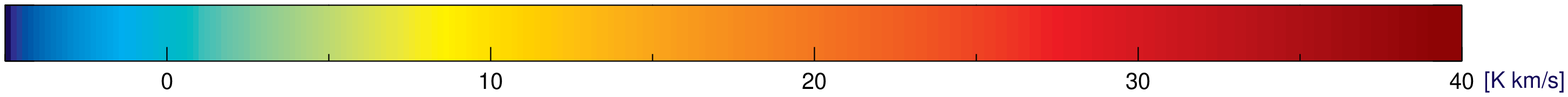}\\
\epsscale{1}
\plotone{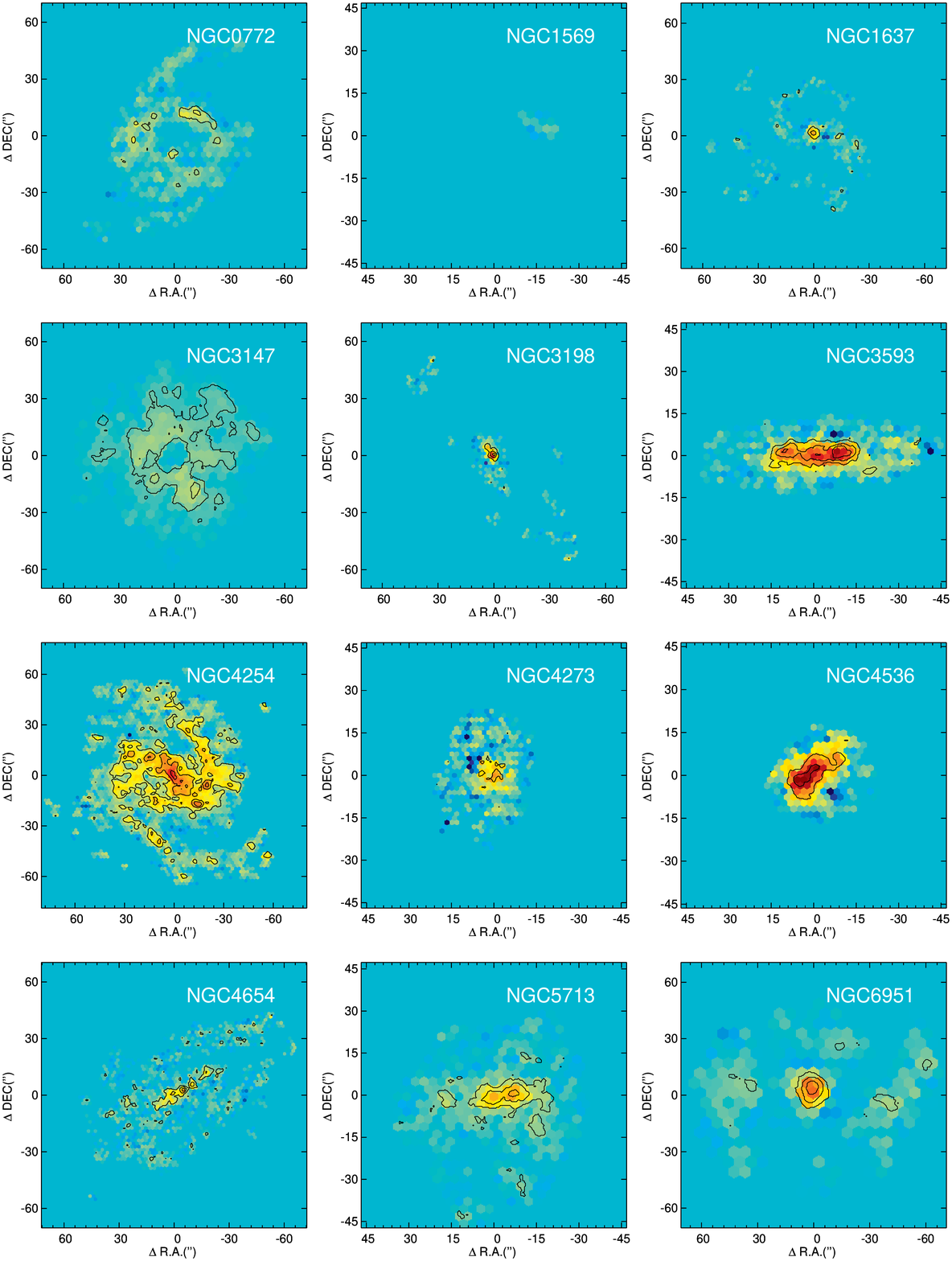}
\caption{\ttline \ integrated intensity maps of the 12 galaxies, 
using same masks as in Figure \ref{fig:12all}. 
Contours show SNR of \ttline  \ at the levels of $[3, 6, 9]$.}
\label{fig:13all}
\end{figure*}

We calculated the velocity-integrated intensity by summing the individual channels from 
the three-dimensional data cube after applying a blanking mask. 
To generate the blanking mask, the \twco \ cube is smoothed 
such that the beam FWHM becomes twice its original value.
Regions with emission greater than $4 \sigma_{\rm ch}$ 
in two adjacent channels of the smoothed cube are used as the core
mask, where $\sigma_{\rm ch}$ is the noise in each individual
channel.
Then the mask is expanded to contain all regions with 
emission greater than $3 \sigma_{\rm ch}$ that are connected
in position or velocity space with the core mask. 
The same \twco \ mask (which we call M12) is used for generating 
both \twco \ and \ttco \ intensity maps for each galaxy, 
based on the assumption that the much weaker  \ttco \ emission should 
be confined to regions where \twco \ is detected.

We generate hexagonal sampling grids with centers
spaced by $\sqrt{3/8}$ times 
the FWHM of the major axis of the synthesized beam,
such that the area of two hexagons equals one beam area approximately.  
Each hexagon is assigned the intensity and other spatially resolved properties 
of the $1 \arcsec$ pixel nearest to its center.
The \twco \ and \ttco \ intensity maps based on such grids 
are shown in Figure \ref{fig:12all} and Figure \ref{fig:13all}. 
Further analysis of intensities throughout this paper 
are based on these hexagonal grids, which we hereafter refer to
as ``half-beams''.
However, fluxes are measured using the original, heavily oversampled
maps (see Sections \ref{sec:spectra} and \ref{sec:global}). 

We divide the half-beams into three groups 
based on their \twco \ and \ttco \ intensities: 
the \ttco-detections have both \twco \ intensity \itw \ and 
\ttco \ intensity \itt \ detected with SNR $>3$,  
the \twco-only-detections have
 \itw \ detected with SNR $>3$ but \itt \ under the detection
threshold of 3, and the \twco-non-detections have 
SNR of \itw \ below 3.  
Note that we use the \twco \ derived mask M12 for obtaining \itw, 
and because \twco \ has higher SNR than \ttco, 
the union of the \ttco-detections and  \twco-only-detections includes all 
of the half-beams that have  \itw \ detected with SNR $> 3$.  

For each of the half-beams, 
the error in integrated intensity was calculated  
by multiplying the square root of the sum of 
$\sigma_{\rm ch}^2$ within the velocity range of the mask 
by the width of the channel.  
The mean $1\sigma$ noise levels of \itw \ (represented as 
$\left<\sigma_{12}\right>$) and 
\ttco \ (represented as $\left<\sigma_{13}\right>$) over the entire maps are listed in Table
\ref{table:mom0}.  
For the \ttco \ integrated intensity, we also calculate the mean $1\sigma$
noise levels of the \ttco-detections 
($\left<\sigma_{13}\right>_{\rm Det}$). 
Note that $\left<\sigma_{13}\right>_{\rm Det}$ is larger than 
$\left<\sigma_{13}\right>$ in each galaxy, 
reflecting the larger velocity width of the M12 mask in
regions where \ttco \ is detected. 
In the last two columns of Table \ref{table:mom0}, we list 
the numbers of half-beams with SNR $>3$ for both \ttco \ and \twco \ respectively.  
The former is the number of the \ttco-detections we defined, and 
the later counts both the \ttco-detections and the 
\twco-only-detections.

\begin{figure*}
\hspace{0.7cm}
\epsscale{1.04}
\plotone{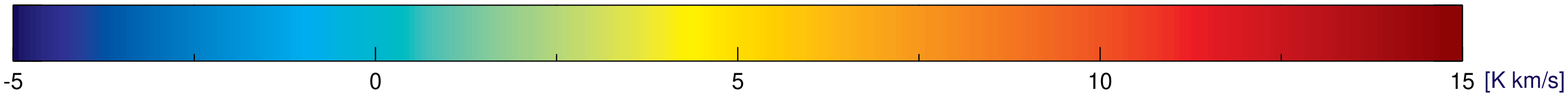}\\
\epsscale{1.0}
\plotone{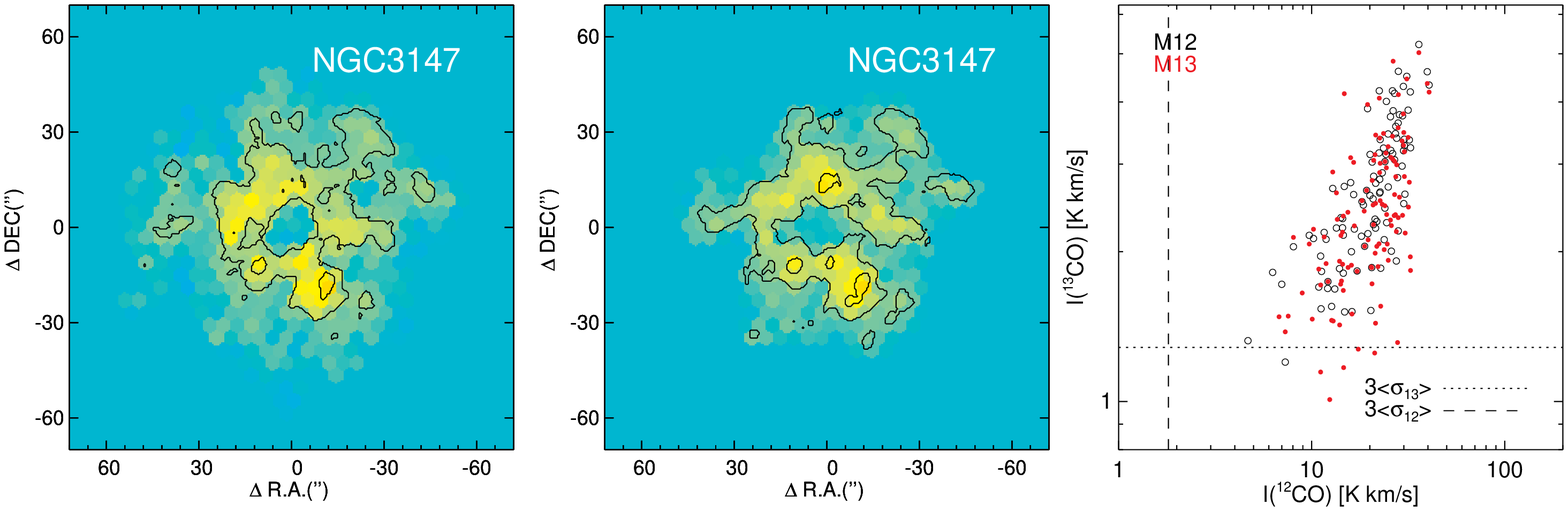}\\
\vspace{-0.3cm}
\plotone{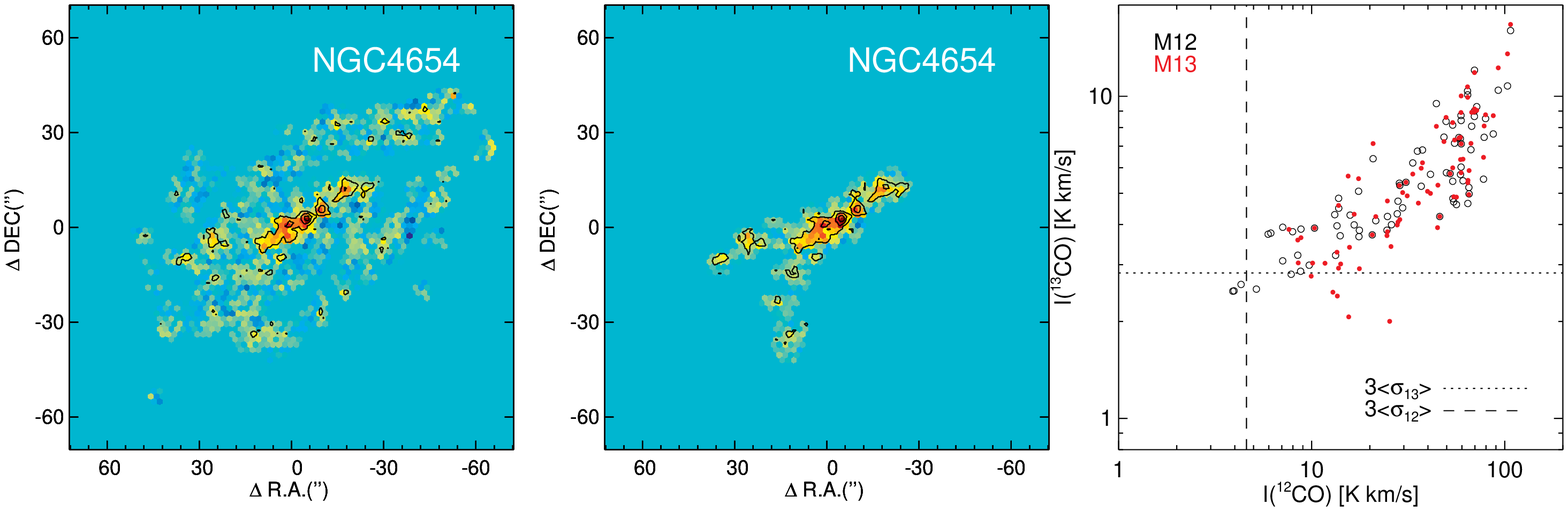}
\caption{Comparisons of \ttco \ maps generated by different masks 
for NGC 3147 and NGC 4654. 
\textit{Left}: \ttco \ integrated intensity map derived with the \twco \ mask (M12);
contours levels are SNR of \ttco \ at [3,6,9].
\textit{Middle}: \ttco \ integrated intensity map derived with the \ttco \ mask
(M13); contours levels are SNR of [3,6,9].
\textit{Right}: Comparison of \ttco \  intensities from M12 and M13 as functions 
of \twco \ intensities. 
Maps from different masks show similar morphologies and distributions 
of \ttco \ intensity.}
\label{fig:13comp}
\end{figure*}

For comparison, we also generated the \ttco \ intensity map 
from an independent blanking mask based on the \ttco \ data cube
(referred as M13) for each galaxy in our sample.
The mask was generated by degrading the angular resolution 
of \ttco \ cube by a factor of $3$, 
and smoothing the spectra by a Gaussian function 
with FWHM of $3$ times the channel width. 
The mask core had a clip level of $4 \sigma$ and 
was extended to adjacent regions or channels 
with emission greater than $2 \sigma$.
These intensity maps were also resampled onto the half-beam hexagonal grids.

Figure \ref{fig:13comp} shows a comparison 
between the two masking methods for NGC 3147 and 4654.
The left panels are the \ttco \ intensity maps 
resulting from the \twco \ masks (M12), while the middle 
panels show the maps derived by applying the \ttco \ masks (M13).
The \ttco \ maps from M12 and M13 have similar
characteristics where \ttco \ emission is strong. 
Some subtle differences result from the
 inclusion of noise from the \ttco \ cube when using the M12 masks; 
the \ttco \ maps generated by applying the M13 masks contain only positive emission, 
whereas the M12 masks sometimes enclose noisy regions
with negative values from the \ttco \ data cubes, 
reducing the integrated intensity. 
In addition, the M12 masks pick up regions with
faint \ttco \ emission extending into the spiral arms 
which cannot be detected using the M13 masks in NGC 4654. 
In the right panels, we show the \ttco \ intensity  
of each half-beam as a function of its \twco \ intensity.
Black circles show intensities resulting from M12, 
while red points show those from M13. 
Three times $\left<\sigma_{13}\right>$ and $\left<\sigma_{12}\right>$ are
shown as the horizontal dotted line and vertical dashed line
respectively.  
Although there are some apparent differences seen in the maps,  
the two masks produce similar distributions of 
\ttco \  for regions having \ttco \ detected with 
SNR greater than 3 (above the dotted line).

\subsection{Flux spectra}\label{sec:spectra}

\begin{figure*}[t!]
\epsscale{1.15}
\plotone{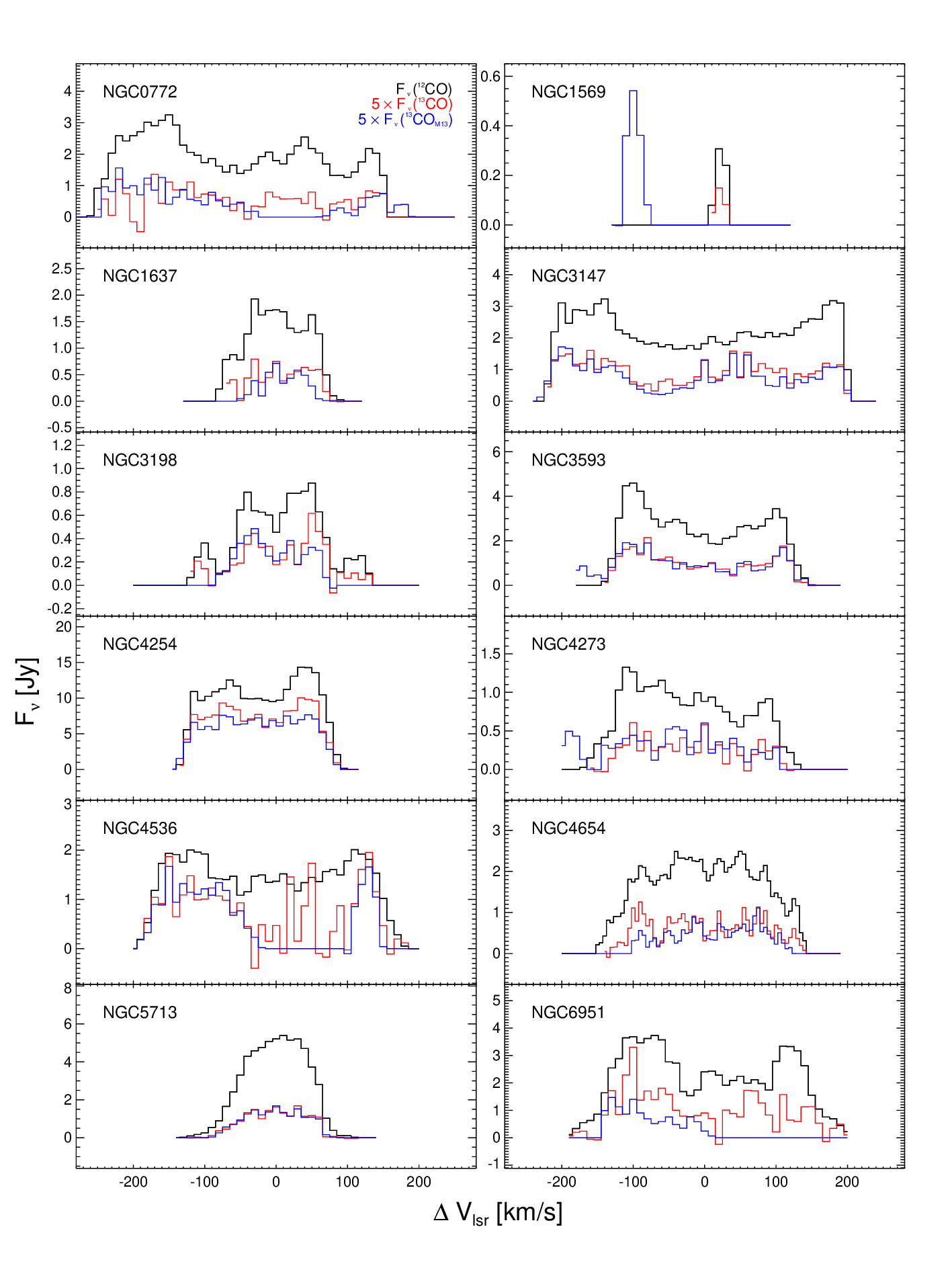}
\caption{
  Total flux spectra plotted as a function of offset from the 
  systemic velocity. Black lines show  the total flux of \twco. Blue
  lines show the \ttco \ flux spectra with the same \twco \ mask (M12), and 
  the red lines are the \ttco \ flux spectra obtained using the \ttco \ mask (M13). 
  The \ttco \ fluxes are scaled by a factor of 5 for clarity. 
}
\label{fig:spectra}
\end{figure*}

For each channel, we sum the line intensity over the extent of the mask 
at the corresponding velocity to obtain the line flux. 
The flux uncertainty per channel is 
calculated by taking the square root of the sum of $\sigma_{\rm ch}^2$ over 
the extent of the mask at that channel. 
Figure \ref{fig:spectra} shows the \twco\ and \ttco\ flux spectra
of each galaxy with the same M12 mask applied, 
as well as the \ttco\ flux spectra with the M13 mask applied for comparison (red histogram).
The \ttco\ spectra are scaled by a factor of 5 for easier viewing.
The \ttco \ emission fluxes derived using the M13 masks are similar to 
those using the M12 masks for channels where the line is detected with high significance.
This confirms the reality of the \ttco\ detections and indicates that 
they are not an artifact of the masking procedure.
On the other hand, the tendency of the \ttco$_{\mathrm M12}$ spectrum 
to recover more flux than the \ttco$_{\mathrm M13}$ spectrum underscores 
the value of using the \twco\ mask as a constraint in cases where the \ttco\ emission is weak.
For NGC 3593 and 4273, there are a few channels with apparent \ttco
\ detections (using the M13 mask) but not corresponding \twco
\ emission.  
Such detections of \ttco \ emission outside the 
M12 mask occur near the edge of the field of view, 
where the primary beam corrections are large, and are thus 
likely to be spurious. 

\begin{deluxetable*}{c r r r r r }
\tablecaption{\label{table:ratio}  Fluxes and line ratios}
\tablehead{
\colhead{Galaxy} &
\mc{$F(\mathrm{^{13}CO})$}  &
\mc{$F(\mathrm{^{13}CO}_{\mathrm M13})$ \tablenotemark{a}} &
\mc{$F(\mathrm{^{12}CO})$} &
\mc{ $\frac{F(\mathrm{^{
          12}CO})}{F(\mathrm{^{13}CO})}$}  & 
\mc{$\left<\mathcal{R}\right>$}
\\
&  
\mc{$(\rm Jy \un km \un s^{-1})$} &
\mc{$(\rm Jy \un km \un s^{-1})$} &
\mc{$(\rm Jy \un km \un s^{-1})$} &
&  
\\
\mc{(1)} &
\mc{(2)} &
\mc{(3)}&
\mc{(4)}&
\mc{(5)}&
\mc{(6)}
}
\startdata
NGC0772  &   $ 45.45 \pm 3.81 $   &   $ 39.31 \pm 2.21 $   &
 $ 816.51 \pm 4.85 $   &   $ 17.97 \pm1.51 $   &   $ 7.63 \pm 2.58 $  \\
NGC1569  &   $ 0.56 \pm 0.39 $   &   $ 2.69 \pm 0.61 $   &   $ 6.26 \pm 0.61 $
  &   $ 11.10 \pm7.71 $   &   \mc{$\ldots$}   \\
NGC1637  &   $ 13.78 \pm 1.62 $   &   $ 8.22 \pm 0.86 $   &
 $ 201.34 \pm 2.54 $   &   $ 14.61 \pm1.73 $   &   $ 11.10 \pm 3.03 $  \\
NGC3147  &   $ 81.21 \pm 3.31 $   &   $ 69.06 \pm 2.52 $   &
 $ 946.14 \pm 4.82 $   &   $ 11.65 \pm0.48 $   &   $ 7.82 \pm 1.65 $  \\
NGC3198  &   $ 11.13 \pm 1.44 $   &   $ 8.62 \pm 1.04 $   &
 $ 112.02 \pm 2.23 $   &   $ 10.06 \pm1.32 $   &   $ 7.03 \pm 2.07 $  \\
NGC3593  &   $ 57.63 \pm 3.21 $   &   $ 62.90 \pm 2.92 $   &
 $ 732.03 \pm 4.93 $   &   $ 12.70 \pm0.71 $   &   $ 8.04 \pm 1.28 $  \\
NGC4254  &   $ 316.19 \pm 4.95 $   &   $ 268.62 \pm 3.76 $   &
 $ 2279.52 \pm 6.83 $   &   $ 7.21 \pm0.11 $   &   $ 5.30 \pm 1.58 $  \\
NGC4273  &   $ 13.97 \pm 1.98 $   &   $ 19.13 \pm 1.50 $   &
 $ 229.05 \pm 3.17 $   &   $ 16.40 \pm2.34 $   &   $ 7.54 \pm 2.31 $  \\
NGC4536  &   $ 54.61 \pm 3.98 $   &   $ 40.43 \pm 2.26 $   &
 $ 525.67 \pm 4.49 $   &   $ 9.63 \pm0.71 $   &   $ 7.56 \pm 2.16 $  \\
NGC4654  &   $ 33.69 \pm 2.01 $   &   $ 21.61 \pm 1.01 $   &
 $ 498.27 \pm 3.40 $   &   $ 14.79 \pm0.89 $   &   $ 7.41 \pm 2.31 $  \\
NGC5713  &   $ 34.05 \pm 2.39 $   &   $ 33.53 \pm 1.83 $   &
 $ 619.37 \pm 3.57 $   &   $ 18.19 \pm1.28 $   &   $ 13.98 \pm 2.90 $  \\
NGC6951  &   $ 73.46 \pm 7.41 $   &   $ 23.57 \pm 2.57 $   &
 $ 828.35 \pm 7.31 $   &   $ 11.28 \pm1.14 $   &   $ 8.65 \pm 1.58 $  \\
\enddata
\tablenotetext{a}{Integrated flux from \ttco \ intensity maps using the M13 mask.}
\end{deluxetable*} 

We present the velocity-integrated fluxes of \ttco \ and \twco \ within the M12 masks 
in columns (2) and (4) in Table \ref{table:ratio}. 
The uncertainties are determined by adding the uncertainties of individual 
channel fluxes in quadrature, and do not take into account uncertainties in defining
the mask.
For comparing the \ttco \ fluxes resulting from different masks, 
we also provide the velocity-integrated \ttco \  fluxes measured in the M13 masks in
column (3) of Table \ref{table:ratio}.  
\ttco \ integrated fluxes using M12 and M13 are similar except for NGC 1569 and NGC 6951. 
As shown in Figure \ref{fig:spectra},
some marginally significant half-beams were detected in M13 but not
M12 for NGC 1569, resulting in more flux using M13 than M12;  
in NGC 6951, low SNR of \ttco \ per channel results in less integrated 
flux measured in the M13 mask than M12. 
We adopt the results obtained with the M12 masks for all subsequent analysis in this 
paper.

\begin{figure*}
\epsscale{1.15}
\plotone{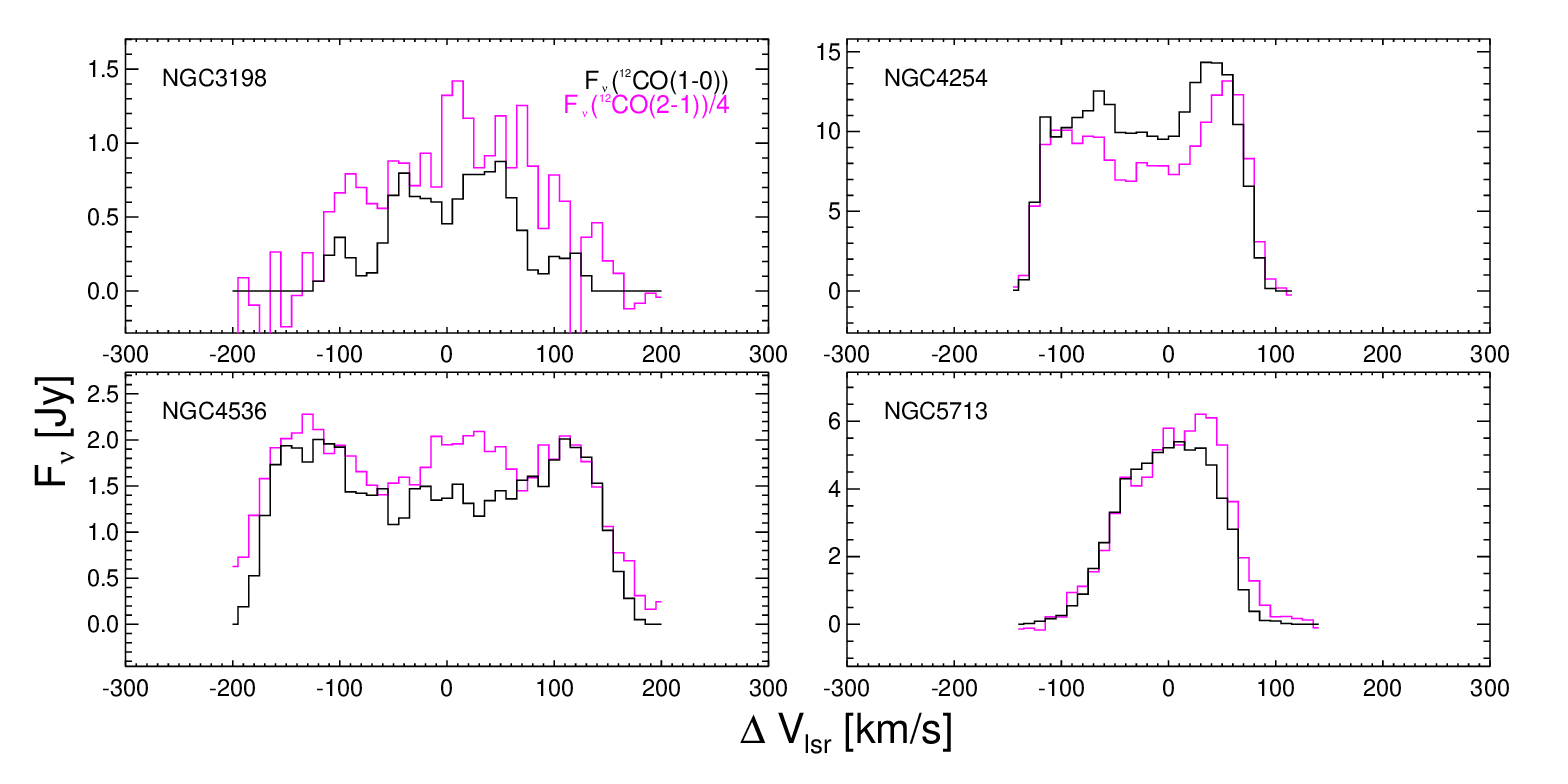}
\caption{
CARMA and IRAM total flux spectra comparison. Black lines show the  total
  flux of \twco$(1 \rightarrow 0)$ using the M12 mask; 
magenta lines show the equivalent \twco($1 \rightarrow 0$) flux
derived from 1/4 of the total flux of \twco($2 \rightarrow 1$) measured by HERACLES.}
\label{fig:spectra_sd}
\end{figure*}

As the interferometer may filter out flux distributed on large scales, 
our data may have underestimated the total flux.
We use single-dish \twco \ $J = 2 \rightarrow 1 $ observations available 
for some galaxies in the sample to compare with our interferometer data.
The single-dish data used are part of 
the HERA CO Line Extragalactic Survey (HERACLES) \citep{Leroy2009} 
using the Institut de Radio Astronomie Millimetrique (IRAM) 
30m telescope. 
The $2.6$ \kms \ velocity channels of the single-dish data were regridded 
to match the interferometer data cube. 
Fluxes measured within a circular aperture of radius $60 \arcsec$ centered on 
the nucleus were used for the comparison. 
The resulting single-dish fluxes were divided by 
a factor of 4 to obtain the equivalent \twco$(J = 1 \rightarrow 0)$ flux,
thus assuming the brightness temperature is approximately 
the same for the \twco$(J = 1 \rightarrow 0)$ and  \twco$(J = 2
\rightarrow 1)$ transitions.

In Figure \ref{fig:spectra_sd}, we show comparisons 
between the masked \twco$(J = 1 \rightarrow 0)$ spectra from 
our interferometer observations and the equivalent \twco$(J = 1
\rightarrow 0)$ spectra
derived from the \twco$(J = 2 \rightarrow 1)$ single-dish data. 
In the four galaxies we compared, 
the interferometer flux spectra have similar characteristics 
to the single-dish spectra. 
Typically, the interferometer recovers more than $80\%$ of the single
dish flux in the majority of channels except for NGC 3198. 
In NGC 3198, the typical ratio of interferometer flux to single 
dish flux is 0.55. In NGC 4536, for a few channels within $\pm 50$
\kms \ of the systemic 
velocity, the flux recovery of CARMA is $10-20\%$ lower than for the
other channels. This is consistent with a more extended distribution
of molecular gas near the minor axis of the galaxy which the 
interferometer is less sensitive to. 
The higher \twco$(J = 2 \rightarrow 1)$ flux could also be due to a steep density gradient or 
higher temperature in the center of NGC 4536. 
We will discuss the possible impact of incomplete flux
recovery on our results in 
Section \ref{sec:diffuse}.  

\subsection{Line ratios}

\subsubsection{Flux ratios} \label{sec:fluxr}

For each galaxy, we calculate a \twco-to-\ttco \ flux ratio,
$F($\twco$)/F($\ttco$)$, by directly taking the ratio of \twco \ and \ttco \ integrated
fluxes in columns (4) and (2) of Table \ref{table:ratio}.  
The resulting flux ratios are presented in column (5) of Table \ref{table:ratio}. 
As the integrated lines fluxes are obtained by summing up all the line emission over the 
extent of the M12 mask, these flux ratios are comparable to ratios of unresolved line 
intensities taken at much lower resolutions by single dishes. 
Our flux ratios range from 7 to 18, with a typical value around 10. 
These values are consistent with flux ratios measured in unresolved studies for 
normal galaxies \citep{Aalto1991, Aalto1995,Eckart1990, S&I1991, Y&S1986, Vila-Vilaro2015}.  
No extremely high values ($\ > 20$) are found in our sample due to the sample selection: 
most of the reported \rtt$\ >20$ values
are from luminous mergers \citep{Casoli1992,
  Aalto1991, Aalto2010}, which are not present in our sample. 
  
 \subsubsection{Intensity ratio maps}
 
\begin{figure*}[ht]
\hspace{0.2cm}
\epsscale{0.98}
\plotone{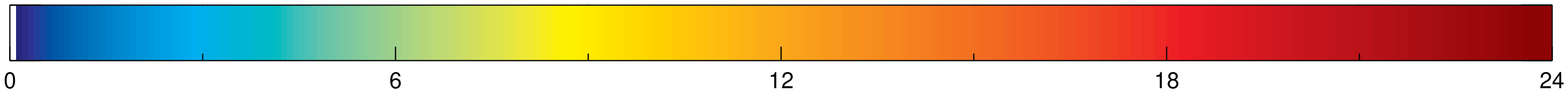}\\
\epsscale{1}
\plotone{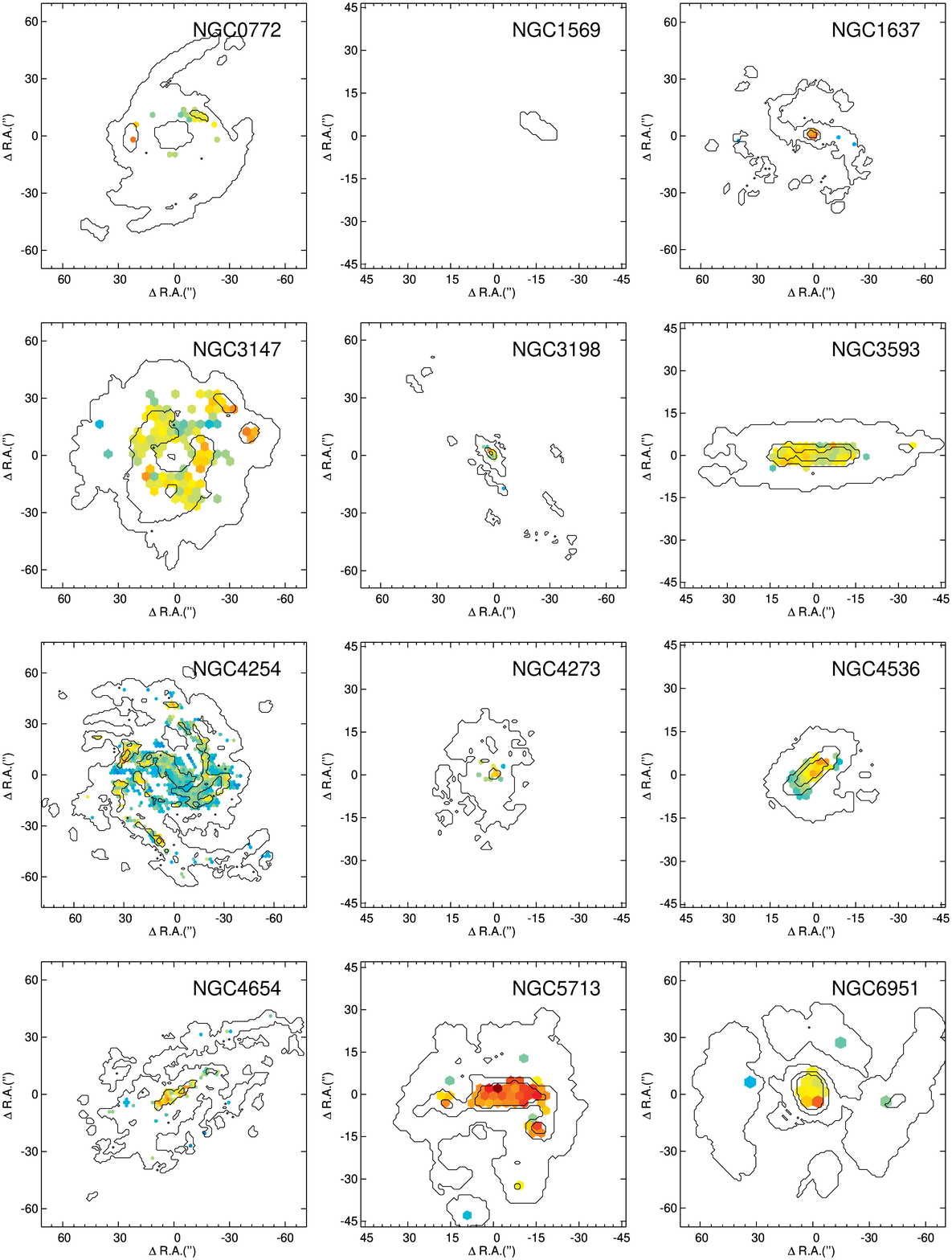}
\caption{
Maps of the \twco \ to \ttco \ intensity ratio, \rtt, 
derived from \ttco \ and \twco \ using the smoothed 
\twco \ mask (M12). 
Only half-beams with \rtt \ detected at $\mathrm{SNR} > 3$ 
are shown here.
Black contours show \twco \ intensity with contour levels of 
$[3, 21, 39]$ times the \twco \ $\sigma_{\rm rms}$.} 
\label{fig:ratio}
\end{figure*}
 
We generated spatially resolved \rtt \ maps for each galaxy from the velocity-integrated intensity maps of \twco \ and
\ttco. 
The uncertainty in \rtt \ at each half-beam was calculated
by propagating the uncertainties in the
\twco \ and \ttco \ intensities.
Figure \ref{fig:ratio} shows maps of the line ratio \rtt, where
measured with $> 3\sigma$ significance, for all of the sample
galaxies. 
The black contours in Figure \ref{fig:ratio} indicate the distribution of \ttco \ emission. 
Within galaxies, \rtt \ in individual half-beams varies by a factor of 3 to 5 
across the disk, mostly ranging from 5 to 15.  
The typical values are consistent with previous \rtt \ measurements of
spiral galaxies on similar scales \citep[e.g.][]{Hirota2010, Pety2013}.
 
Within our detection limit, comparing to studies with similar sensitivities, we do not have reliable \rtt \ measurements higher than $30$ as
reported by some previous studies near starbursting regions 
\citep{Huttemeister2000, Aalto2010}.    
Nor did we find dramatic variations of \rtt \ within each galaxy,
while spatial variations in \rtt \ of more than a factor of 3  
over scales of $ 1 \un kpc$ are seen in some starbursting galaxies
\citep{Huttemeister2000, Aalto2010, Meier2004}.
The rather moderate \rtt \ values and their limited variations  
are partly because our galaxy sample does not include luminous mergers or
starbursting galaxies,  in which higher \rtt \ and
strong spatial variations in physical properties have been reported.
On the other hand, our limited sensitivity to \ttco  \ emission  
confines the reliable \rtt \ measurements to regions with stronger \itt, 
while regions with high \rtt \ but weaker \itt \ might not appear in the \rtt \ map.  
Moreover, our typical resolution of several hundred parsecs may smear out
abnormal values of \rtt, and also make it difficult to identify   
rapid changes in \rtt \  or gas properties that occur on
scales of less than $100\un pc$ \citep{Meier2004}.   

\subsubsection{Weighted mean ratios} 
Based on the resolved \rtt \ maps,
we computed a mean value for the intensity ratio in a galaxy, $\left<\mathcal{R}\right>$, 
by taking the mean of resolved \rtt \ for the \ttco-detections    
weighted by their corresponding \ttco \ intensities:   
\beq
\left<\mathcal{R}\right> = \frac{\sum\limits_{i=1}^{\rm Det}
  I_{13,i}\mathcal{R}_{i}}{\sum\limits_{i=1}^{\rm Det} I_{13,i}}
= \frac{\sum\limits_{i=1}^{\rm Det} I_{12,i}}{\sum\limits_{i=1}^{\rm Det} I_{13,i}}
\eeq
where $i$ is a half-beam in the set of \ttco-detections. 
The \ttco \ intensity weighted mean \rtt, $\left<\mathcal{R}\right>$, is thus
the ratio of \twco \ and \ttco \ fluxes for the \ttco-detections.  
The $\left<\mathcal{R}\right>$ value  
and the weighted standard deviation for each galaxy  
are listed in column (6) of Table \ref{table:ratio}.  
These $\left<\mathcal{R}\right>$ values range between $5.3 \pm 1.58$ for NGC 4254
and $13.98 \pm 2.9$ for NGC 5713, with a typical value of $\sim$ 7--9. 
These values are similar to typical line ratios measured on (sub)-kpc scales 
\citep[e.g.][]{Paglione2001, Hirota2010, Wilson1994}.

\begin{figure}
\epsscale{1.2}
\plotone{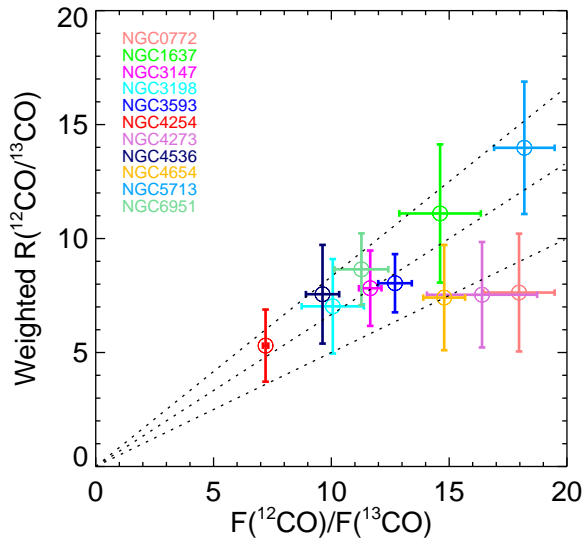}
\caption{Weighted mean \rtt \ as a function of integrated flux ratio 
$F($\twco$)/F($\ttco$)$. Each colored point represents a galaxy from STING. 
The vertical error bars show the weighted standard deviation of \rtt, 
and the horizontal error bars show the uncertainty of $F($\twco$)/F($\ttco$)$. 
The dashed lines show scaling factors of 1, 2 and 3 from the top to
the bottom.}
\label{fig:rvsfr}
\end{figure}

In Figure \ref{fig:rvsfr}, we show  
a comparison between the weighted mean ratio
$\left<\mathcal{R}\right>$ 
and the flux ratio $F($\twco$)/F($\ttco$)$ obtained 
in Section \ref{sec:fluxr}.
The flux ratio of a galaxy is larger than 
the weighted mean $\left<\mathcal{R}\right>$,
because total flux ratios are calculated over all 
regions within the aperture, 
while $\left<\mathcal{R}\right>$ 
is the flux ratio for the \ttco-detections which  
have brighter \ttco \ emission and are likely biased to lower \rtt \ ratios. 
We consider this bias in greater detail in the following section.

\subsection{Detection bias} \label{sec:bias}

 \begin{figure*}
\epsscale{0.95}
\plotone{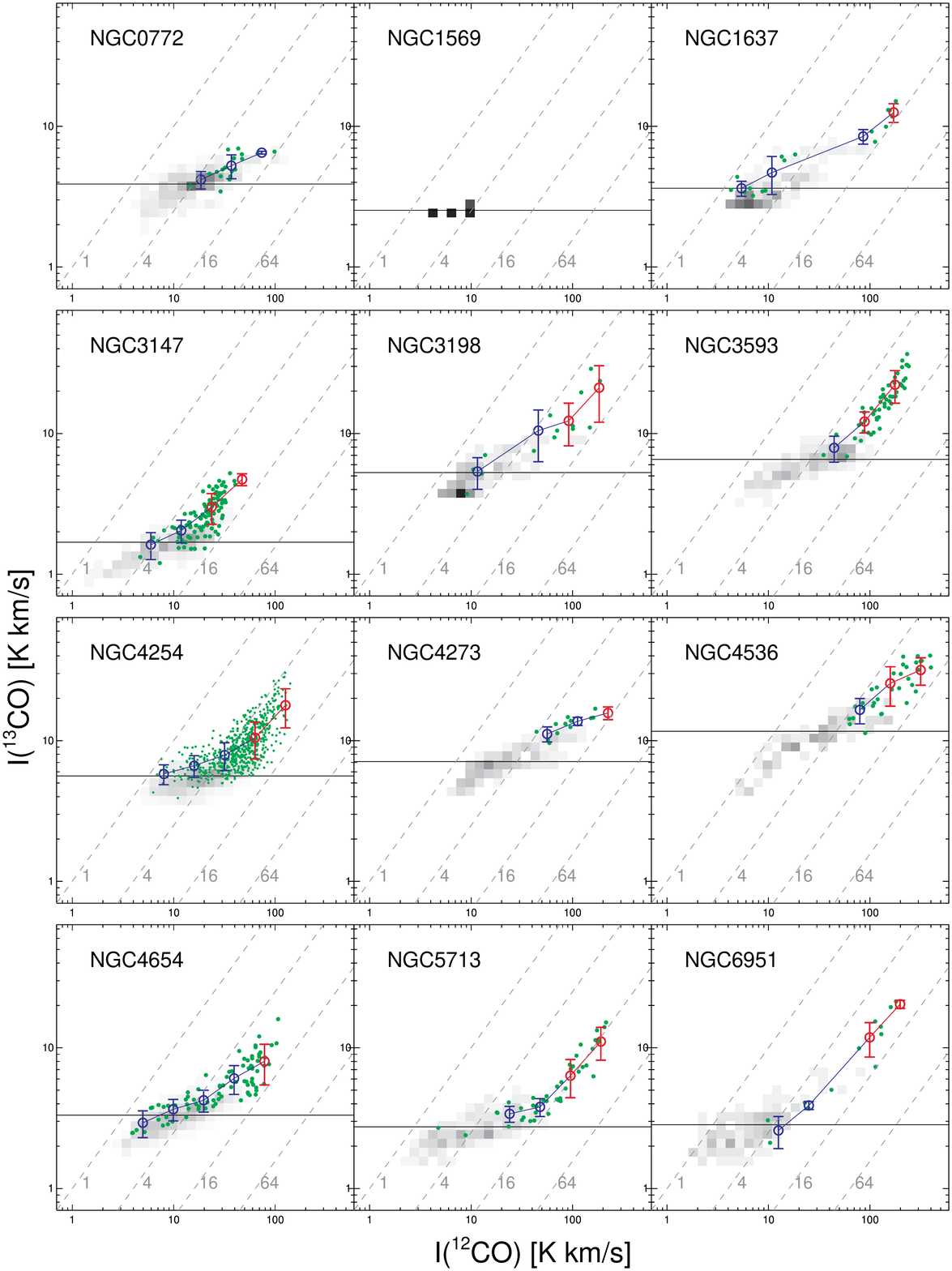}
\vspace{-1cm}
\caption{
Intensity of the \ttco \  emission as a function of the
  \twco \ emission of each galaxy. 
Green dots are the half-beams with \ttco \ detections.   
Gray scales show the distribution of upper limits of \ttco \  
in the regions with \twco \ SNR$ > 3$ but without detectable \ttco \  emission. 
All intensities are sampled onto half-beams defined in Section \ref{sec:mom0}. 
The linked circles show the mean \itt\ value for \ttco \ detections in each \itw \ bin, 
with red symbols representing the bins in which the number of \ttco \ detections is more than 
$50\%$ of all the half-beams.
The error bars on the circles show the $1\sigma$ scatter.  
The horizontal solid line is the typical threshold of $3\sigma_{\rm 13}$, and the dashed lines show constant \rtt \ of 1, 4, 16 and 64.   
}
\label{fig:i13vs12}
\end{figure*}

In Figure \ref{fig:ratio}, the resolved \rtt \ measurements are 
available for much fewer regions than those with \twco-detections,
leaving large numbers of half-beams within the \twco \ contours blank. 
For a half-beam to be measured in \rtt, the typical SNR of \ttco \ needs to be larger than 4;
this threshold will select \twco-detections with SNR(\itw) $> 3$\rtt, 
as the typical $\sigma_{12} \sim 1.2 \sigma_{13}$.
Given LTE assumptions with $T_k = 10 \un K$, [\twco]/[\ttco]$=40$,
and \rtt$\sim 8$, the \twco-detections only require 
$N_{\mathrm{H_2}} >  3 \times 10^{20}\, \mathrm{cm}^{-2}$,
while the \rtt \ measurements require column densities 8 times larger.
Moreover, the detection threshold also implies that the measured \rtt \ has 
an upper limit of SNR(\itw)/3, which can be quite restrictive in cases of low SNR. 
For a sample consisting mostly of \twco-detections without \rtt \ detections,
the resolved \rtt \ are biased to lower values and 
should not be used to infer the mean or median \rtt \ of the sample.

To quantify how the bias of \rtt \  due to the limited detection threshold 
could affect our data, we show values of \itw \ and \itt \ of the resolved \rtt \  
measurements for each of the galaxies in Figure \ref{fig:i13vs12}.
Solid dots show individual half-beams with \rtt \ measurements,
and open circles show the mean value in each bin.
Error bars on the circles show the $1\sigma$  scatter of the data.
We further divide these mean values in two categories
according to the detection fractions of the corresponding bin,
defined as the ratio of the number of \rtt \ detections to the number of \twco-detections.
Red symbols represent those with detection fraction greater than $50\%$,
and blue ones show the bins with detection fraction below $50\%$.
The horizontal solid line in each panel shows
the detection threshold set by $3\sigma_{\rm 13}$;
for simplicity, we use the typical value of $\sigma_{\rm 13}$ of the galaxy from column 3 of Table \ref{table:mom0}. 
The gray shaded region show the upper limits of \itt \ inferred from $3\sigma_{\rm 13}$ for those half-beams below the detection threshold.
For reference, we plot constant \rtt \ values in the figure as dashed lines;
\rtt \  increases from the top left to the bottom right.

For the half-beams with resolved measurements,
there appears a clear trend that the \rtt \ are lower at the fainter end of \itw.
However, the upper limit of \rtt, set by the detection threshold  of $3\sigma_{\rm 13}$,
also increases with \itw \ as SNR of \itw \  increases; 
the lower \rtt \ measured in the low \itw \  regime could
result from a selection effect as higher \rtt \ values are not measurable. 

To derive an estimate of \rtt \ for the bins with a large fraction of 
half-beams below the detection threshold,  we need to improve the SNR of \ttco.  
In the following, we will first use spectral stacking to estimate
the average \ttco \ intensity for these bins.  
A toy model is then developed to provide one possible
explanation of the distribution of \itt \ shown in our data.

\subsubsection{Stacked line ratios}
\label{sec:stack}

\begin{figure*}
\epsscale{0.95}
\plotone{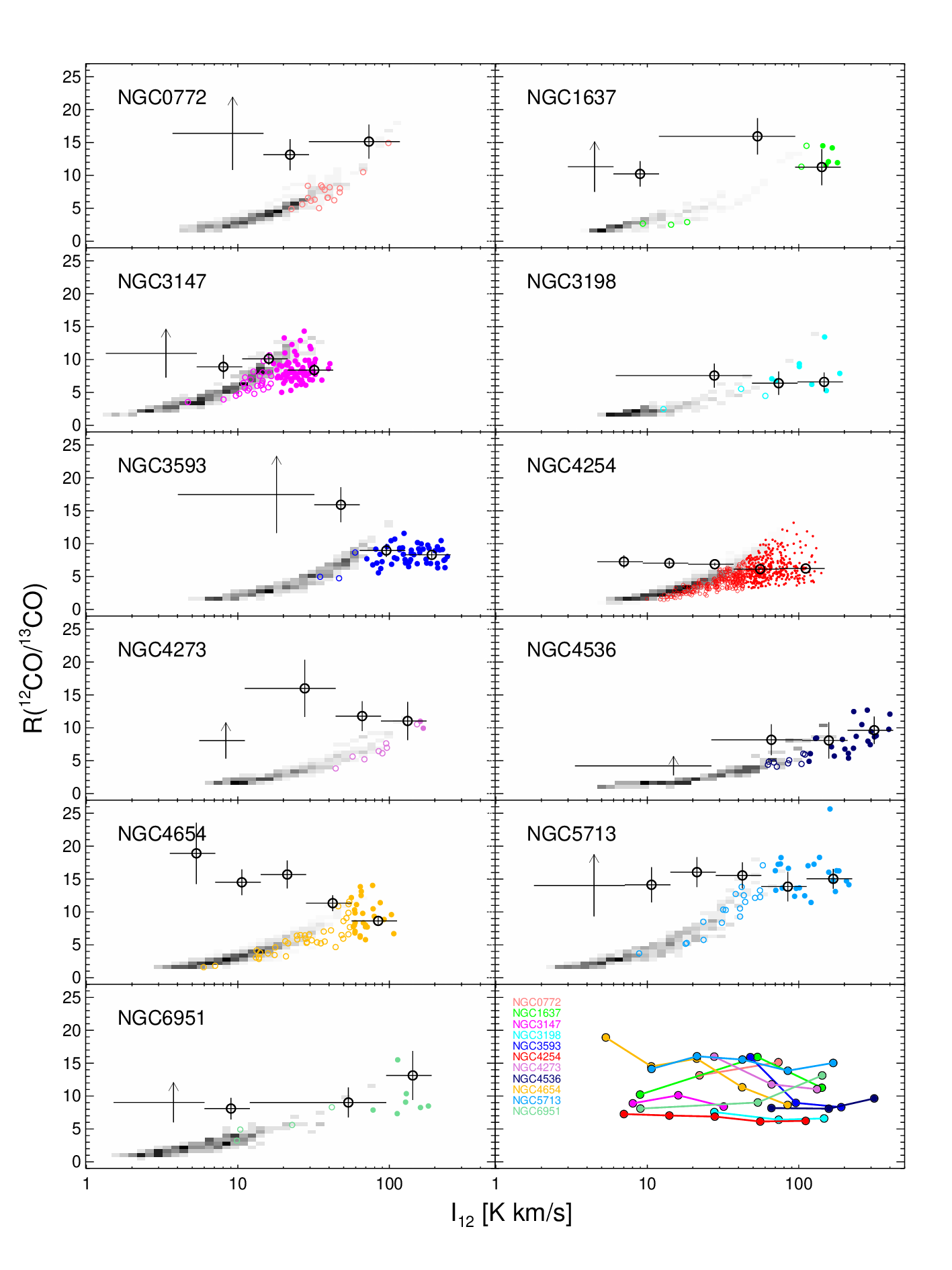}
\caption{
Resolved  and stacked \rtt \ as a function of \itw. 
The first 11 panels show the results for each individual galaxy, excluding NGC 1569. 
Small colored circles show the ratios for individual half-beams 
with \rtt \ detected; the filled ones show the half-beams in \itw \ bins with 
detection fraction greater than $50\%$, and the open ones show the half-beams 
in \itw \ bins with lower detection fraction. 
The gray scales show the distribution of 
\rtt$_{\rm min} = $\itw$/3\sigma_{13}$ for half-beams with \twco \ detected 
but \rtt \ not detected.  
The black large circles and arrows are the stacked \rtt \ and lower limits 
in each \itw bin respectively; 
the horizontal error bars reflect the bin size, and the vertical error bars 
show their uncertainties. 
The last panel is a summary plot showing stacked \rtt \ 
with different colors representing different galaxies.
}
\label{fig:r13vsi12}
\end{figure*}

To compensate for the relatively low SNR of \ttco \ which introduces a  detection bias on resolved \rtt, we stack spectra of a number of half-beams to estimate their average line intensities and line ratios. 
By stacking spectra for $N$ beams, we are able to increase  the SNR of \ttco \ by a factor of $\sqrt{N}$ 
and expand our analysis to a larger dynamic range.
Only \twco-detections are used for stacking as their velocities can be well measured from \twco. 
The \twco \ and \ttco \  spectra of each half-beam are shifted to a common velocity using its \itw-weighted mean velocity, and stacked with others based on physical properties such as galactic radii and SFRs.     
We fit both stacked spectral lines by Gaussian profiles 
to derive the integrated intensities. We use profile fitting because it does not require a predefined signal window for integration, and it can better extract the line emission rather than baseline structure.  
For each stacked \ttco \ spectrum detected with peak SNR above 3, 
the stacked line ratio is calculated by taking the ratio of stacked integrated \twco \ and \ttco \ intensities. 
When the stacked \ttco \ intensity's peak SNR is below 3, the uncertainty of integrated \ttco \ is calculated by integrating the channel noise over the FWHM of the stacked \twco.  We use 3 times this uncertainty as the upper limit of \ttco \ to derive the lower limit of stacked \rtt.

We first investigate the stacked line ratio as a function of \itw. 
We use logarithmic bins of \itw \ with a step of 0.3 dex. 
If two adjacent bins both have stacked \itw \ with SNR $< 3$, we merge them into one. Bin merging continues,  in the direction of decreasing SNR, until the lower limit of \rtt \ exceeds 10, the stacked line ratio has SNR $>3$, or there are no more bins to merge. 
These stacked line ratios or their lower limits as a function of \itw \ are shown 
in Figure \ref{fig:r13vsi12}.
The black circles show stacked \rtt \ in each bin of \twco. 
The stacked \rtt \  does not always increase with \itw, unlike the case for the resolved \rtt  \ of \ttco-detections.
In the bins with weaker \itw \ and a large fraction of \ttco \ non-detections 
(corresponding lower limits are shown in gray scale),      
stacked \rtt \ are higher than the individual  \ttco-detections; 
for these bins the \ttco-detections are likely a biased sample from  the overall population 
in the bin. 
The apparent increasing trend of resolved \rtt \ with \itw \ 
seen in Figure \ref{fig:i13vs12} is 
mainly due to this detection bias in the lower \itw \ bins.

On the other hand, for the bins with stronger \itw \ and high detection fraction,  resolved \rtt \ are less biased and the stacked ratios are similar to the mean values. 
Filled color dots in each panel show the resolved \rtt \ in bins for which detection fraction is more than $50\%$ (the bins shown in red symbols in Figure \ref{fig:i13vs12}).    
For these half-beams, the resolved \rtt \ is less biased and does not strongly depend on \itw. 

\subsubsection{A toy model} \label{sec:biasmd}

In nearby galaxies, \itw \ is measured on a much larger scale than a 
single GMC, so that brighter \itw \ in a beam also means more clouds are
included; the scatter in \rtt \ could decrease with \itw \ by averaging more clouds in a telescope beam. If all the resolved \rtt \ detected consist of GMCs with similar properties and \rtt, it is possible that the dependence of resolved \rtt \ on \itw \ results from applying detection thresholds on \rtt \ with decreasing scatter.  We have developed a simple model as described below to test this possibility. 

We divide the molecular gas in a galaxy into small scale parcels
(``cloudlets'') each emitting the same amount of \twco \ (\itwc), 
so the \itw \ we measure in a half-beam is a proxy for the
number of cloudlets in it. 
We assume that the \ttco \ emission of each cloudlet (\ittc) in the galaxy 
is drawn from the same probability distribution and is 
independent of other cloudlets. 
We use \itwc \ and \ittc \ 
to represent the \twco \ and \ttco \ intensities of a single cloudlet,
as opposed to other quantities without an upper index 
representing measurements over a half-beam.  
The inverse of \rtt \ is then the average of
\ittc/\itwc \  
over all the cloudlets in a half-beam:
\beq
\frac{1}{\mathcal{R}} = 
\frac{\sum\limits_{i=1}^{N} I_{13}^{\rm cl}}{N I_{12}^{\rm cl}} 
= \frac{\left<I_{13}^{\rm cl}\right>}{I_{12}^{\rm cl}}
= \left<\frac{1}{\mathcal{R}^{\rm cl}}\right> ,
\eeq
where \rtt \ is the line ratio for a half-beam, $N$ is the number of cloudlets in the half-beam, and $\mathcal{R}^{\rm cl}$ is the ratio of a single cloudlet.
For independent and identically distributed $1/\mathcal{R}^{\rm cl}$ 
(\ittc/\itwc), $1/\mathcal{R}$ of a half-beam, as the average value of $1/\mathcal{R}^{\rm cl}$ for the $N$ cloudlets in it, will have 
the same expectation as $1/\mathcal{R}^{\rm cl}$ and  remain constant throughout the galaxy. 
Meanwhile, the variance of $1/$\rtt \ for a half-beam will be 
the variance of $1/\mathcal{R}^{\rm cl}$ \ divided by the number of cloudlets $N$ the half-beam contains. 
To implement the model and compare it with the data,
we logarithmically binned all the half-beams with \twco \ 
SNR $> 3$ detections according to their \itw. 
We choose the bin with the highest \twco \ intensity as the reference 
bin, because \ttco \ measurements in that bin are the least 
biased by sensitivity.    
We measured the mean and variance of $1/$\rtt \ in the reference bin 
as $I_{13,\mathrm{ref}}/I_{12,\rm ref}$($1/$\rtt$_{\mathrm{ref}}$) and $S_{\rm ref}^2$. 
For half-beams with a given \itw, the expectation and variance 
of $1/$\rtt \ are:  
\beq
E\left[\frac{1}{\mathcal{R}}\right]  = \frac{I_{13,\mathrm{ref}}}{I_{12,\rm ref}} = \frac{1}{\mathcal{R}_{\mathrm{ref}}},
\eeq
\beq
 \mathrm{Var}\left[\frac{1}{\mathcal{R}}\right]= S_{\mathrm{ref}}^2 \frac{I_{\mathrm{12, ref}}}{I_{12}}.
\eeq

Moreover, by the central limit theorem,  
\itt/\itw \ of a half-beam, as the mean of a number of 
independent $1/$\rtt$_{\mathrm{cl}}$,  
will tend to be normally distributed if the half-beam contains 
a sufficiently large number of cloudlets. 
Besides the intrinsic scatter of $1/$\rtt \ assumed by the toy model, we also include measurement uncertainties using typical values of $\sigma_{\rm 13}$ of the galaxy from column 3 of Table \ref{table:mom0} into the toy model.
In our approach, we model the half-beams in NGC 4254 only if they
have \itw \ above a lower limit of $20 \rm \  K \ km \ s^{-1}$, 
which corresponds to approximately 500 cloudlets each 
with a typical mass of $10^4 \ M_{\odot}$ using a standard CO-to-H$_{2}$ conversion factor of 
$X_{\rm CO} = 2 \times 10^{20} \ \mathrm{cm^{-2} (K \ km
  \ s^{-1})^{-1}}$.      
Aside from the deviation from a normal distribution due to the small  
number of cloudlets,  
for half-beams with \itw \ under the limit of 
$20 \rm\  K \ km \ s^{-1}$, 
measurement errors in \itw \  introduce
an additional source of scatter in $1/$\rtt \  that is not included in the model.  
Therefore, we only model the distribution of $1/$\rtt \ 
for each single half-beam with $I_{12} > 20 \rm \  K \ km \ s^{-1}$, 
including both \ttco \ detections and non-detections.  

\begin{figure*}
\epsscale{0.75}
\plotone{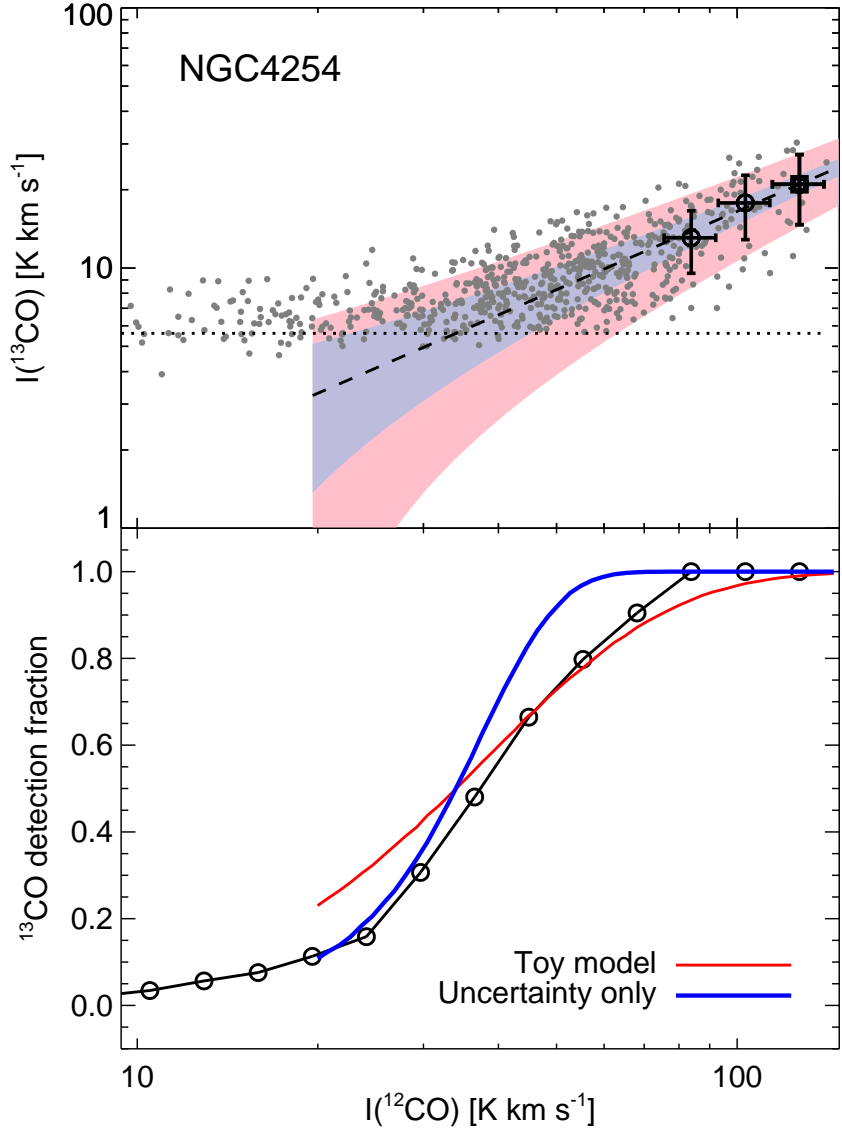}
\caption{
Comparison of the observed \itt \ and its detection fraction in NGC 4254  with a toy model. 
\textit{Upper panel}: \itt \ as a function of $I_{\mathrm 12}$. 
The gray points show half-beams with \itt \ detected 
above $3\sigma_{\rm 13}$.  
The black square symbol shows the mean \itt \ of the reference bin chosen for measuring $1/\mathcal{R}_{\rm ref}$ and $S_{\rm ref}$, and the black filled circles with error bars show \itt \ of the other 2 bins in which all the half-beams are detected in both \ttco \ and \twco; horizontal and vertical error bars on these symbols show the width and standard deviation of each bin respectively.   
The dashed line shows the constant \rtt \ assumed by the model.
The dotted black line shows the detection
threshold set by 3 times the mean sensitivity of \ttco.  
The red shaded band shows the range of standard deviation of the toy model, and the blue shaded band shows the measurement uncertainty of \itt \ assuming a constant \rtt. 
\textit{Lower panel}: Detection fraction of \ttco \ as a
function of  $I_{\mathrm  12}$. 
The circles show detection fractions of \ttco \ with SNR $>3$ from observation. 
The red line shows the detection fractions
predicted by the toy model, and the blue line shows the detection fractions obtained by assuming the scatter on \rtt \ is entirely from the measurement uncertainty.}
\label{fig:mdr}
\end{figure*}

In Figure \ref{fig:mdr}, we show the results of such an approach 
for NGC 4254 for illustration.
NGC 4254 has the largest number of \ttco-detections in our sample of
galaxies, providing the most data points for estimating
the statistical properties of the reference bin
($1/$\rtt\ and $S_{\rm ref}$) 
used in the model.
The upper panel shows \itt \  as a function of \itw.
The gray points are the measured \itt\  detected where both 
\twco \ and \ttco \ have SNR $> 3$.  
The black square symbol shows the reference \itw \ bin ($I_{12, \rm
  ref}$) in which we measure  $I_{13,\mathrm{ref}}/I_{12,\rm ref}$ and $S_{\rm ref}$,  
with the vertical error bar showing the standard deviation of \itt \  and the 
horizontal error bar showing the \itw \ bin size.
The other 2 bins with all the half-beams  detected in \ttco \ are shown as black filled circles. All these bins show similar mean \rtt \ to the constant mean value of \rtt \  
indicated by the dashed line in the plot.
We use a black dotted line to show the detection threshold of $3\sigma_{13}$. The red shaded band spans the 1$\sigma$ distribution of the
\itt \ we expect in the model. By comparison, we use a blue shaded band to show the 1$\sigma$ distribution \ of \itt \ when only including the measurement uncertainty into the constant \rtt \ without intrinsic scatter. 
In the high \itw \ regime, 
the distribution of the detected \itt \ looks very similar to the prediction of the toy model.
In the low \itw \ regime, the detected \itt \ values are located at the upper 
end of the modeled distribution. 
Although individually these detections seem to deviate from the
average \itt \ of the model shown as the dashed line, 
their distribution is still close to the 1$\sigma$ envelope of the modeled distribution.   
In contrast, the measurement uncertainty shown in the blue band is smaller than the observed scatter measured in the three bins with highest \itw, and many of the individual detections with lower \itw \ are also outside of the envelope. 
Including measurement uncertainty only does not reproduce the 
observed \itt \ distribution very well.

For each \itw \ bin, we further compare the 
observed \ttco \ detection fraction
with the fraction of \ttco \ measurements predicted by the model to be 
above the detection threshold. 
The black circles in the lower panel are the detection fractions 
from the data with SNR $> 3$. 
The red solid line shows the detection fraction 
of the model by setting a detection threshold of
$3\sigma_{13}$, 
i.e.\ the fraction of modeled \ttco \ above the 
dotted line in the upper panel. 
The detection fraction calculated for constant \rtt \ with measurement uncertainties is shown in blue.
As the figure shows, the toy model predicts very similar detection fractions 
to the observations, while a model which only includes the measurement uncertainty shows detection fractions less favored by the data. 
The success of the toy model in predicting the detection fraction 
over a factor of 10 in \itw \ suggests that 
dependence of resolved \rtt \ on \itw \  can result primarily from  
a combination of intrinsic scatter and detection bias.

\section{Results} \label{sec:results}

\subsection{Line ratio as a function of galaxy properties} \label{sec:global}

The STING sample spans a range of galaxy properties,
such as stellar mass, star-formation rate, and morphology.
Before examining the spatially resolved \rtt \ in each
galaxy, we first compare the global line ratios among the sample galaxies,
to study the dependence of 
line ratios on galaxy properties. 
For each property, we first investigate the flux ratio   $F($\twco$)/F($\ttco$)$, which 
 is determined by the gas content of the entire galaxy. 
We then study the \ttco \ intensity weighted mean ratio 
$\left<\mathcal{R}\right>$ that represents the typical value of the resolved \rtt \ on (sub-)kpc scales.  

\subsubsection{Dust temperature}

\begin{figure*}
\plottwo{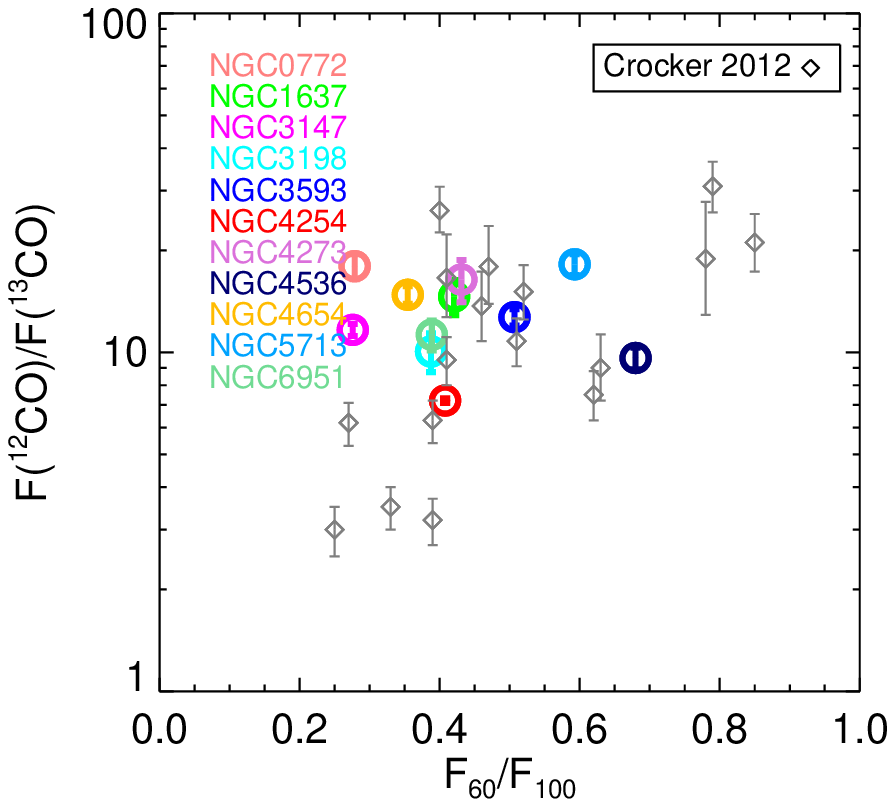}{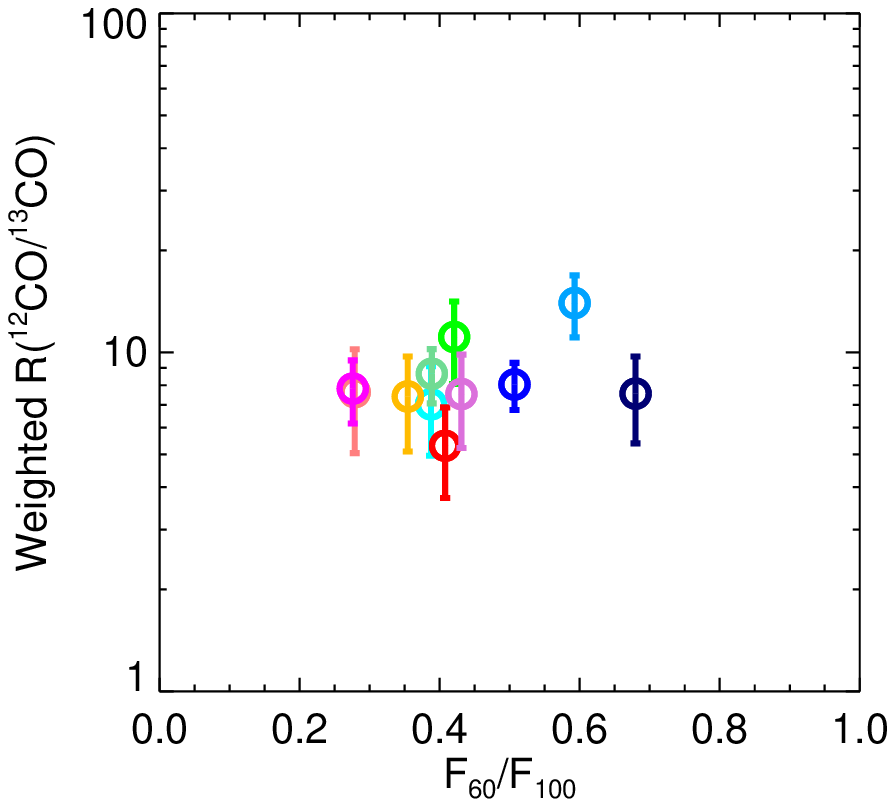}
\caption{
\textit{Left}: Flux ratio $F($\twco$)/F($\ttco$)$ vs. the far-IR flux 
ratio at 60 and 100 $\rm \mu m$ ($F_{60}/F_{100}$), with the 
color circles representing the STING galaxies and the 
black diamonds showing samples from \citet{Crocker2012}. 
For the STING galaxies, 
the vertical error bars
show the uncertainty in the flux ratio. 
\textit{Right}: \ttco \ intensity weighted mean ratio 
$\left<\mathcal{R}\right>$  vs. $F_{60}/F_{100}$. 
The vertical error bars
show the \ttco \  intensity weighted standard deviation of \rtt. 
}
\label{fig:rvsf61}
\end{figure*}

We plot the flux ratio $F($\twco$)/F($\ttco$)$ as a function 
of \textit{IRAS} IR colors $F_{60}/F_{100}$ in the left panel of Figure \ref{fig:rvsf61},
together with the data from \citet{Crocker2012}.
For $F_{60}/F_{100}$ between 0.2 and 0.6, we do not find 
a strong correlation between $F($\twco$)/F($\ttco$)$ and $F_{60}/F_{100}$.
In previous studies, although a positive correlation between
the two parameters has been claimed,
for  $F_{60}/F_{100} \lesssim  0.6$, the correlation 
is tentative and shows large scatter
\citep{Y&S1986,Aalto1995, S&I1991,  Crocker2012}. 
A recent survey of \ttco \ in normal galaxies by \citet{Vila-Vilaro2015} 
also found no correlation between $F($\twco$)/F($\ttco$)$ and $F_{60}/F_{100}$.  
Our results are in agreement with these studies. 

The $F_{60}/F_{100}$ ratio is often considered as an indicator of
dust temperature. A higher dust temperature might be associated 
with a higher gas temperature, which would reduce the average opacity of \twco \ and increase  $F($\twco$)/F($\ttco$)$.  
A higher dust temperature also implies the galaxy has more active star formation, which may increase the fraction 
of gas in a diffuse molecular phase 
(with low opacity and high \rtt) through feedback. 
There is also the possibility that chemical fractionation
towards \ttco \ at lower dust temperatures results in
low \rtt \ at low $F_{60}/F_{100}$ \citep{Crocker2012}. 
All of these effects could lead to a positive correlation between  
$F($\twco$)/F($\ttco$)$ and dust temperature.  
However, our results show no such correlation for  $F_{60}/F_{100} \sim 0.2 - 0.6$; 
the effect of dust temperature on the line ratio is not prominent.  

The right panel of Figure \ref{fig:rvsf61} shows that 
the \ttco \ intensity weighted mean ratio 
$\left<\mathcal{R}\right>$ also shows little dependence on $F_{60}/F_{100}$.  
For a galaxy with higher dust temperature and thus higher star formation rate on average, 
systematically lower optical depth of \twco \ in molecular clouds might result from 
star formation heating the gas and/or broadening the line width. 
Therefore, a positive correlation is expected if such opacity effects 
determine the resolved \rtt. 
Our results suggest that systematic differences in resolved \rtt \ cannot be attributed to different dust temperatures.

\subsubsection{Inclination}

\begin{figure*}
\plottwo{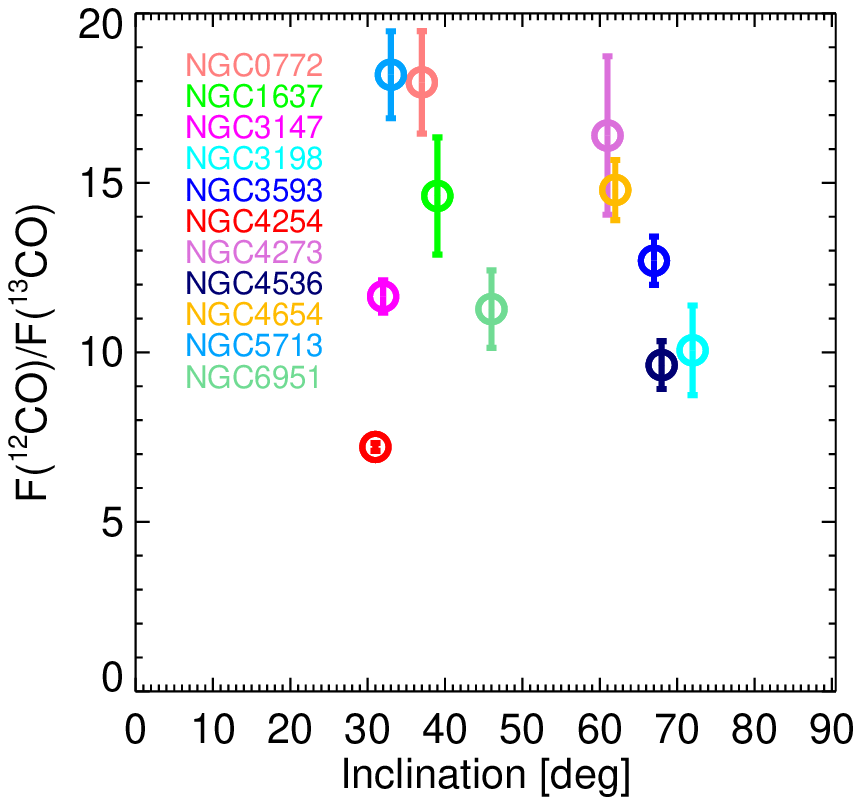}{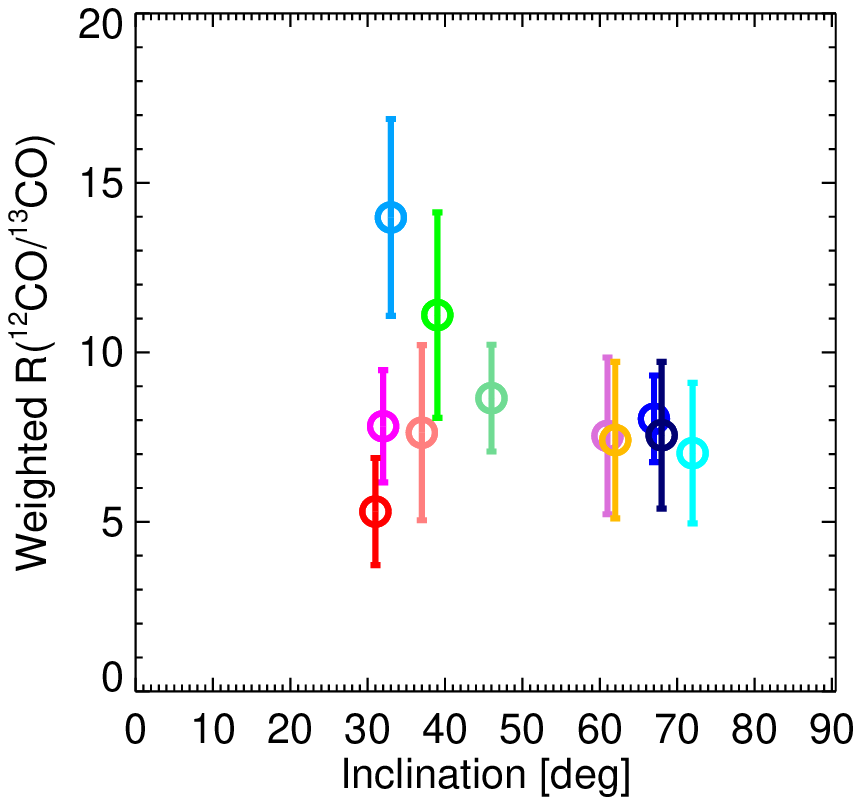} 
\caption{
\textit{Left}: Flux ratio $F($\twco$)/F($\ttco$)$ 
as a function of galaxy inclination. 
The vertical error bars 
show the uncertainty in the flux ratio.
\textit{Right}: \ttco \ intensity weighted mean ratio 
$\left<\mathcal{R}\right>$ as a function of 
galaxy inclination. The vertical error bars 
show the intensity weighted standard deviation of \rtt.
}
\label{fig:rvsinc}
\end{figure*}

The galaxy inclination may affect line ratios by changing the column density along the line-of-sight. 
Highly inclined galaxies will have larger average optical depth than face-on galaxies, 
and thus lower $F($\twco$)/F($\ttco$)$, 
unless line broadening due to rotation reduces the optical depth per velocity channel. 
However, there is no such negative  correlation between 
$F($\twco$)/F($\ttco$)$ and galaxy inclination, as shown in the left panel of Figure \ref{fig:rvsinc}. 
The results are consistent with previous findings by  \citet{Y&S1986} and \citet{S&I1991}, 
which also suggest   that inclination has no effect on the large-scale line ratio. 

For the resolved \rtt, the number of molecular clouds within a beam will increase with the inclination.  
Therefore, \rtt \ in more inclined galaxies will be measured by averaging
more clouds than in face-on galaxies, 
leading to less variations of \rtt \ within the galaxy. 
We present the weighted mean ratio $\left<\mathcal{R}\right>$ as
a function of inclination in the right panel of Figure \ref{fig:rvsinc}. 
The typical resolved \rtt \ of a galaxy does not show a dependence on the galaxy's inclination; 
there is also no strong effect of inclination on (sub-)kpc scales. 
The error bars show the standard deviation of resolved \rtt, 
and we do not find a correlation between the standard deviation of \rtt \ and the inclination. 
This differs from the finding by \citet{S&I1991} who found that 
the variance of the line ratio is larger in face-on galaxies.    
However, note that the standard deviation of \rtt \ we calculate includes 
only the \ttco-detections, which might underestimate the spatial variance of \rtt \ in the galaxy (c.f. Figure \ref{fig:mdr}).  

\subsubsection{Metallicity}

\begin{figure*}
\plottwo{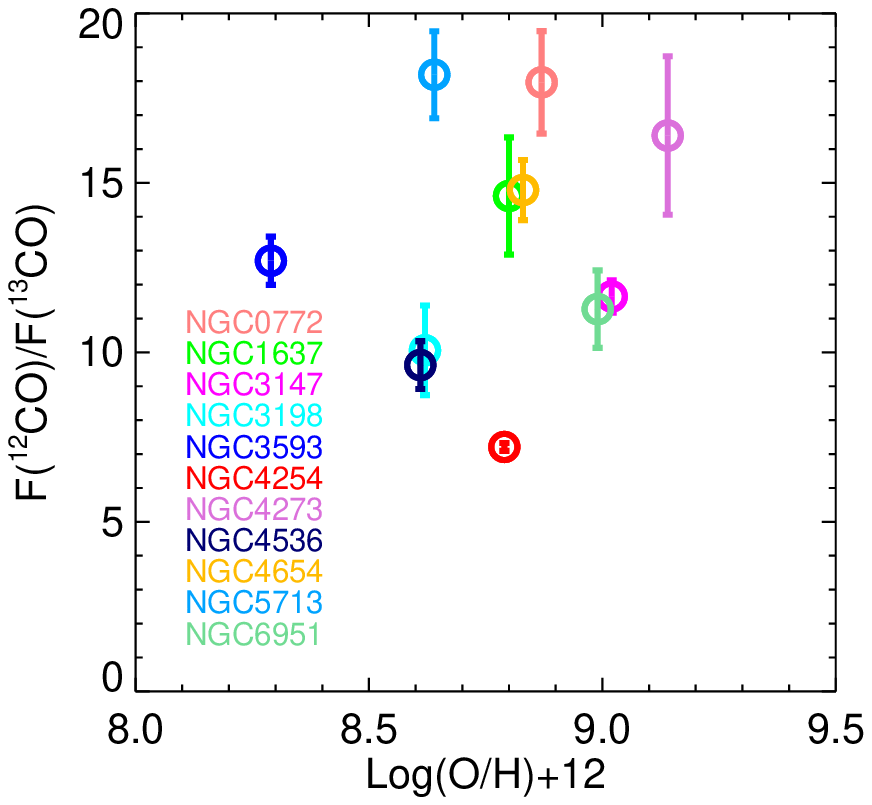}{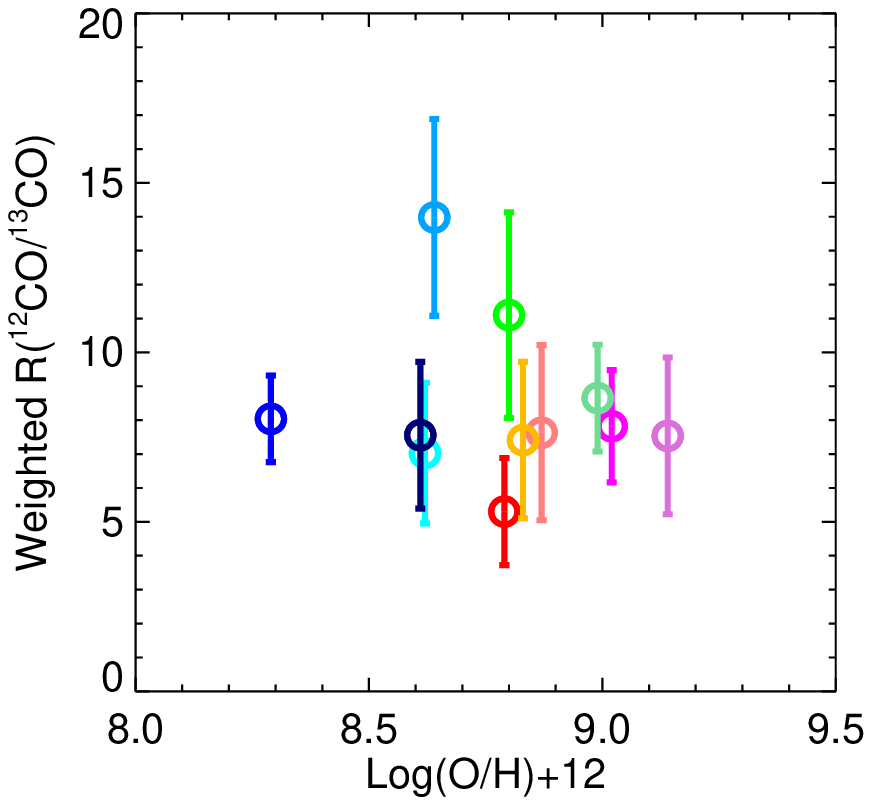} 
\caption{
\textit{Left}: Flux ratio $F($\twco$)/F($\ttco$)$ 
as a function of galaxy metallicity. 
The vertical error bars 
show the uncertainty in the flux ratio. 
The references for  
metallicity are listed in Table \ref{table:ginfo}.
\textit{Right}: \ttco \ intensity weighted mean ratio 
$\left<\mathcal{R}\right>$ as a function of 
galaxy metallicity. The vertical error bars 
show the intensity weighted standard deviation of \rtt.
}
\label{fig:rvsmt}
\end{figure*}

In the left panel of Figure \ref{fig:rvsmt}, we show the flux ratio $F($\twco$)/F($\ttco$)$ as a function of gas-phase metallicity. 
Although our sample spans a considerable range in metallicity, 
there is no clear evidence that $F($\twco$)/F($\ttco$)$ is directly correlated with metallicity.
This is in agreement with previous studies using larger samples 
\citep{S&I1991, Crocker2012}. 
Theoretically, an anti-correlation between $F($\twco$)/F($\ttco$)$ and metallicity is expected: in metal-poor galaxies, because \twco \ should be less abundant, 
the average optical depth of \twco \ emission will be reduced and
thus $F($\twco$)/F($\ttco$)$ will be elevated. 
However, this expectation assumes a single-component LTE model, which is probably too simplistic to apply to a galaxy as a whole. 
\ttco \ and \twco \ very likely
originate from different structures in the line of sight:
\ttco \ should emit from a deeper layer than the well-shielded \twco.
In low metallicity environments,
\twco \ is less effectively shielded so its emitting area shrinks \citep[e.g.][]{Wolfire2010}, while
\ttco \ in denser regions might be less affected.
Therefore, the increase of the flux ratio $F($\twco$)/F($\ttco$)$ for a 
metal-poor galaxy may be less than expected.

Metallicity could also affect the line ratios through changes in $[^{12}\rm C]/[^{13}\rm C]$. 
As $^{13}\rm C$ is produced mainly by intermediate-mass stars 
and  accumulates slowly as a galaxy evolves, 
in low metallicity environments where gas is less processed,     
under-abundant $^{13}\rm C$ and hence a higher fractional abundance
$[^{12}\rm C]/[^{13}\rm C]$ will also tend to increase $F($\twco$)/F($\ttco$)$ 
as well as the resolved \rtt. 
$F($\twco$)/F($\ttco$)$  and the weighted mean ratio $\left<\mathcal{R}\right>$ are 
both expected to anti-correlate with metallicity 
if $[^{12}\rm C]/[^{13}\rm C]$ decreases with increasing metallicity. 
However, as shown in Figure \ref{fig:rvsmt}, 
$F($\twco$)/F($\ttco$)$ or $\left<\mathcal{R}\right>$ does not depend on metallicity in our sample. 
There are several possible reasons for this discrepancy between the
observed results and these expectations.
First of all, because there are only a few direct measurements
of $[^{12}\rm C]/[^{13}\rm C]$ reported,
the relation between the fractional abundance and metallicity remains unclear. 
One unanticipated finding was that $[^{12}\rm C/^{13}\rm C]$ of  
clouds in the Small and the Large Magellanic Clouds is 
similar to that of Milky Way clouds \citep{Heikkila1999}.
In addition, the impact of $[^{12}\rm C/^{13}\rm C]$ on  $F($\twco$)/F($\ttco$)$ and $\left<\mathcal{R}\right>$ 
could still be very limited or even washed-out by the other factors
that change the opacity, such as has been found in the Milky Way: 
while $[^{12}\rm C]/[^{13}\rm C]$ has a positive radial gradient in distance from the Galactic center \citep{Milam2005}, \rtt \ varies from $\sim 10$ in the Galactic center \citep{Bally1987,Martin2004} to $\sim 6$ in Galactocentric radius range of $3-6 \rm \ kpc$ \citep{Roman-Duval2016}

\subsubsection{Significance of trends}

\begin{figure*}
\plottwo{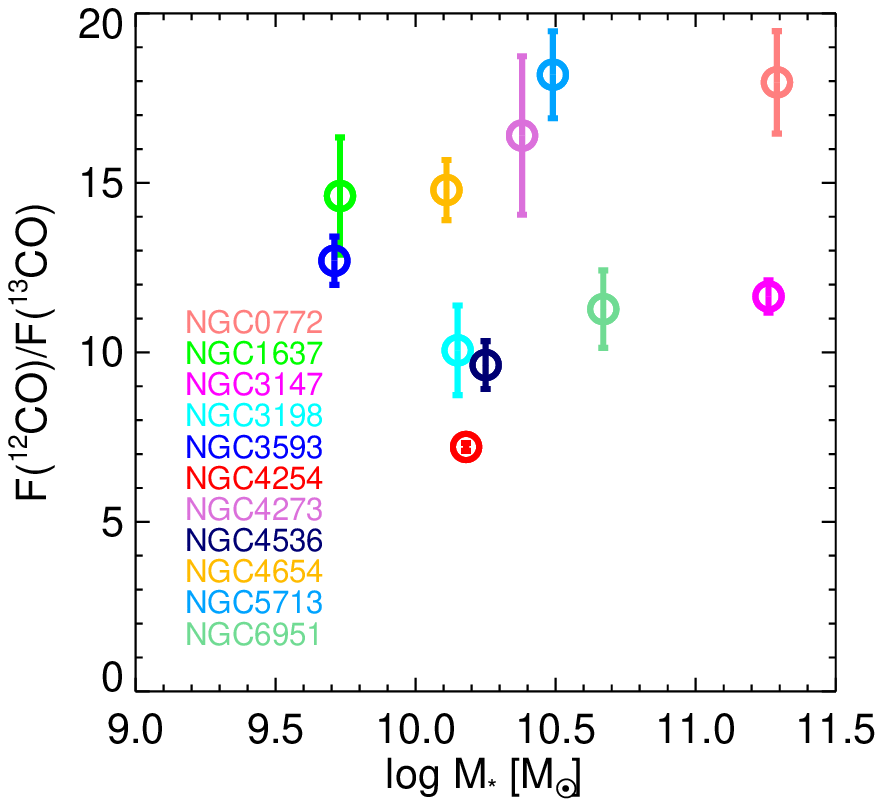}{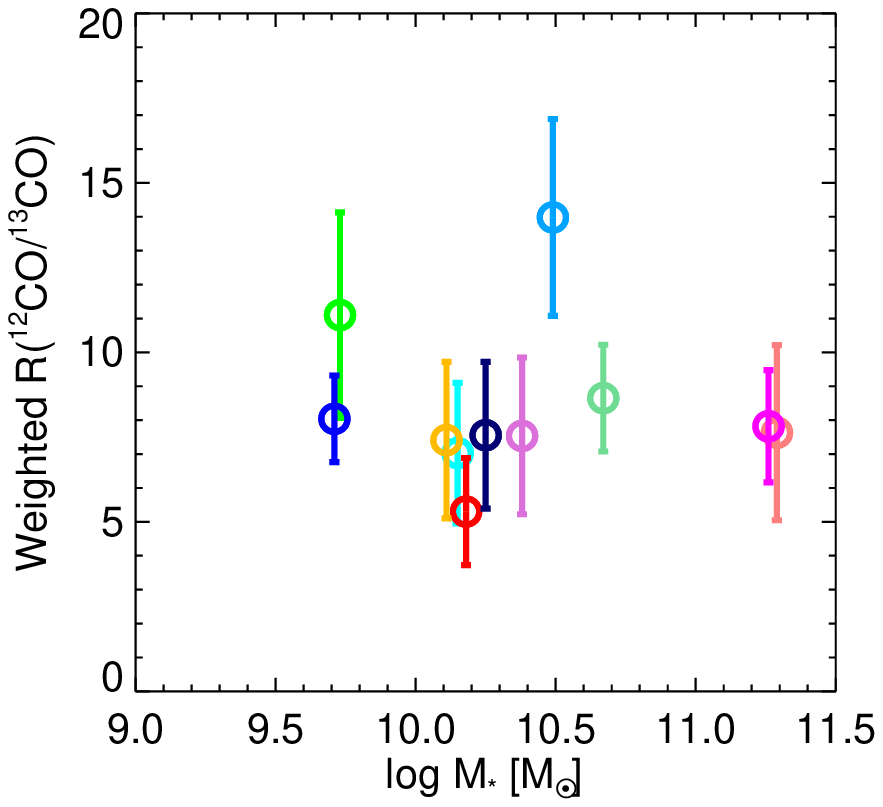} 
\caption{
\textit{Left}: Flux ratio $F($\twco$)/F($\ttco$)$ 
as a function of stellar mass of the galaxy. 
The vertical error bars 
show the uncertainty in the flux ratio. 
\textit{Right}: \ttco \ intensity weighted mean ratio 
$\left<\mathcal{R}\right>$ as a function of 
stellar mass. The vertical error bars 
show the intensity weighted standard deviation of \rtt.
}
\label{fig:rvsms}
\end{figure*}

\begin{deluxetable}{l r c c r c}
\tablewidth{0pt}
\tablecaption{\label{table:global} Line ratio and galaxy properties}
\tablehead{
\colhead{Galaxy property} &  
 \multicolumn{2}{c}{ $F($\twco$)/F($\ttco$)$} & &
 \multicolumn{2}{c}{$\left<\mathcal{R}\right>$} \\
 \cline{2-3}   \cline{5-6}  
 &
\mc{$r_{\rm s} $\tablenotemark{a}} &
\mc{$P_{0}$ \tablenotemark{b}} & &
\mc{$r_{\rm s} $  \tablenotemark{c}} &
\mc{$P_{0}$ \tablenotemark{d}}
}

\startdata
$F_{60}/F_{100}$\tablenotemark{e} & 0.27 & 0.42 &  & $-$0.02 & 0.96\\
Inclination & $-$0.17 & 0.61 & &  $-$0.22 & 0.52 \\
$\log{\rm O/H} + 12$ &  0.32&  0.34& & 0.01& 0.98 \\
$\log{M_{*}}$\tablenotemark{f} & 0.07 & 0.84 & & 0.09 & 0.80 \\
\enddata

\tablenotetext{a}{Spearman's rank correlation coefficients between $F($\twco$)/F($\ttco$)$ and galaxy properties.}
\tablenotetext{b}{Probability of the null hypothesis of no correlation between  $F($\twco$)/F($\ttco$)$ and galaxy properties.}
\tablenotetext{c}{Spearman's rank correlation coefficients between  $\left<\mathcal{R}\right>$ and galaxy properties. }
\tablenotetext{d}{Probability of the null hypothesis of no correlation between  $\left<\mathcal{R}\right>$ and galaxy properties.}
\tablenotetext{e}{The $60 \rm \ \mu m$ and $100 \rm \ \mu m$ fluxes are from NED. }
\tablenotetext{f}{Stellar masses are derived from from $3.6 \rm \mu m $ luminosity using Equation C1 in \citet{Leroy2008}.} 
  
\end{deluxetable}

In Table \ref{table:global}, we list the
Spearman rank correlation coefficients ($r_{\rm s}$) between the line ratios 
and each galaxy property we have investigated, and the significance
of its deviation from zero ($P_{0}$).
The significance $P_{0}$ represents
the probability assigned to the hypothesis that the variables are
unrelated; a lower $P_{0}$ means a more significant correlation. 
For our sample of 11 galaxies, we expect a significant correlation when $|r_s| > 0.8$ and $P_{0} < 0.05$.
The results shown in Table \ref{table:global} further confirm the lack of strong trends 
between \twco/\ttco\ line ratios the galaxy's dust temperature, inclination, and metallicity. 
In addition to these three properties, we also tested for a correlation between
the line ratio and the stellar mass of the galaxy. The flux ratio and the weighted mean ratio as functions of stellar mass are shown in Figure \ref{fig:rvsms}.  No strong trend between \rtt \ and stellar mass was found in our sample.

\subsection{Variations of spatially resolved line ratio} \label{sec:spatial}

In this section, we investigate the variations of resolved \rtt \ 
within each galaxy as measured in half-beam width apertures. 
Both of the stacked and resolved \rtt \ are analyzed. 
The stacked intensities and line ratios are derived following the same approach described in Section \ref{sec:stack}.  
The stacked values are used for studying the mean trends.
Since we restrict the stacking to places where resolved \twco \ is well detected, 
the stacked spectra will tend to overestimate the mean \twco \ and yield a higher line ratio 
if the bin also includes significant number of \twco \ non-detections. Therefore, we 
excluded the bins in which \twco-detections number less than \twco-non-detections.
For quantifying the resolved correlations between \rtt \ and other local properties, we only use the half-beams \ in bright \itw \ bins for which detection fraction is more than $50\%$ (the bins shown in red symbols in Figure \ref{fig:i13vs12}), since these resolved \rtt \ do not strongly depend on \itw \ and are less affected by the sensitivity bias.

\subsubsection{Radial dependence}

\begin{figure*}[ht]
\plotone{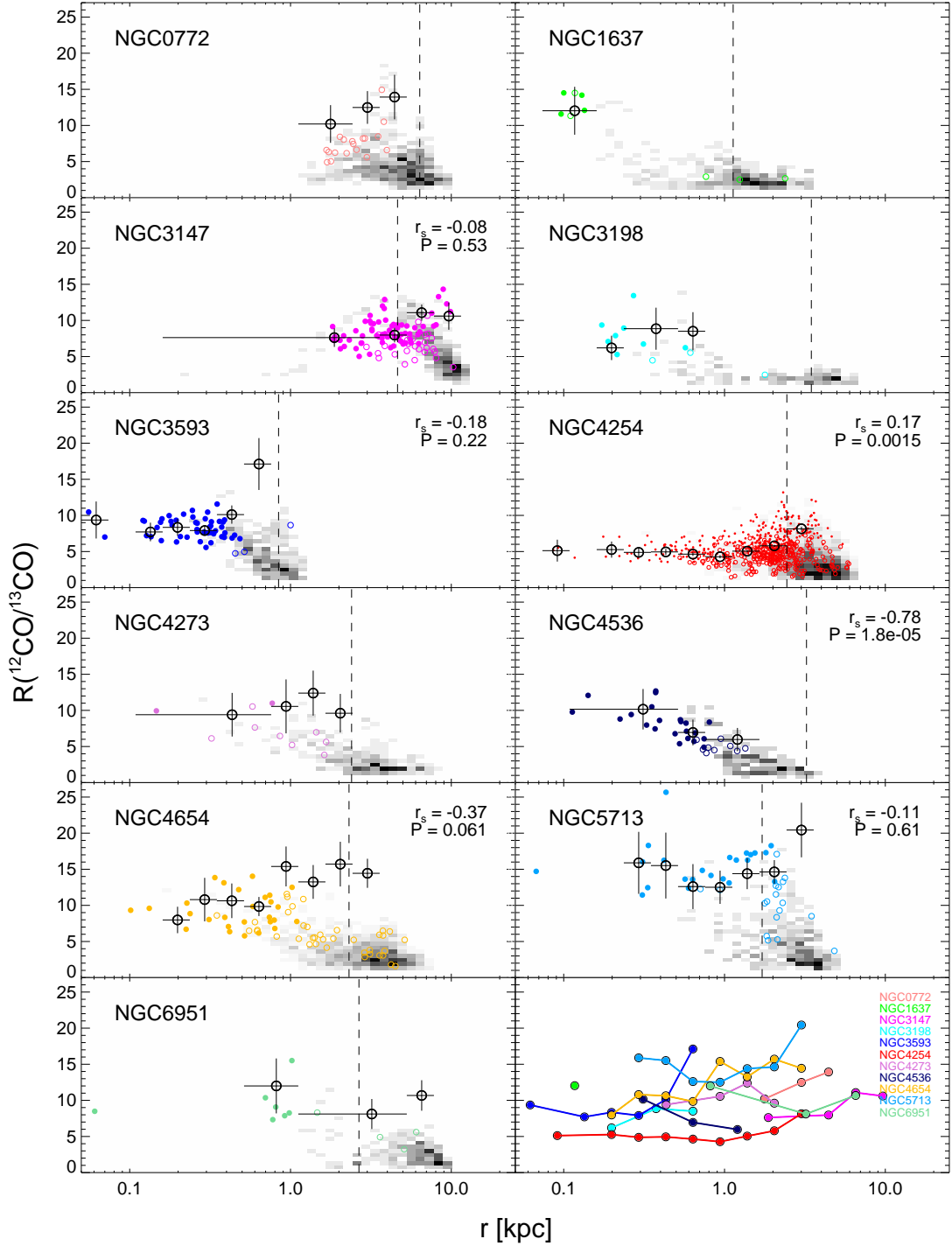}
\caption{
Line intensity ratio \rtt \ as a function of radius. 
The first 11 panels show the results for each individual galaxy.
Small colored circles show the ratios for individual half-beams 
with \rtt \ detected, with the filled ones highlight those 
from which the correlation coefficients are derived. 
The gray scales show the distribution of 
\rtt$_{\rm min} = $\itw$/3\sigma_{13}$ for those with \twco \ detected 
but \rtt \ not detected.  
The black large circles are the stacked \rtt \ in each radius bin; 
the horizontal error bars reflect the bin size, and the vertical error bars 
show their uncertainties. 
The vertical dotted lines show the positions of $0.2 R_{25}$.
Spearman's rank correlation coefficient $r_s$ and 
the significance $P$ are shown in the top right of panels with $>$20 valid half-beams. 
The last panel is a summary plot showing stacked \rtt, 
with different colors representing different galaxies.
}
\label{fig:rradius}
\end{figure*}

The radial distribution of \rtt \ has been studied in 
previous works, but their findings do not offer a consistent  picture. 
In the Milky Way, \citet{Roman-Duval2016} found that \rtt \  increases with galactocentric distance.
While \citet{R&B1985} and \citet{Paglione2001} claimed \rtt \ is 
generally higher in the central regions than the disks 
in their surveys, there are a handful of galaxies 
showing constant radial profiles \citep{Y&S1986, S&I1991}, 
and others showing \rtt \ increasing with radius \citep[e.g. M51,][]{Pety2013}.

Figure \ref{fig:rradius} shows the resolved and stacked \rtt \ 
as functions of galactocentric radii. 
The resolved \rtt \  are shown in the plot by individual colored points, 
while the distribution of lower limits for non-detections are shown in gray shading.
We binned the galactocentric radii into equal bins in log space and adjusted them 
until \rtt \ in each stacked bin has SNR$>3$ or the stacked lower limit exceeds 10. 
The open circles show stacked \rtt \ in each bin. 
The colored dots show the \rtt \ detections.  
Filled color dots in each panel show the half-beams in bright \itw \ bins used to measure the correlation coefficients, 
and we require more than 20 such half-beams in a galaxy for the measurement. 
We show Spearman's correlation coefficient $r_s$ and 
the significance level ($P_{0}$) of the null hypothesis 
in the top right of each panel.  
We expect a significant correlation when $|r_s| > 0.6$ and $P_{0} < 0.05$.
The gray shading shows the lower limits of \rtt \ derived by using upper limits of \itt \ for the non-detections.

Of 6 galaxies with correlation coefficients available, 
only NGC 4536 exhibits a clear tend with radius, in the sense that 
 \rtt \ decreases with radius. 
Such radial trends are often found in starbursting galaxies
\citep[e.g.][]{Aalto2010, Tosaki2002, Hirota2010}. 
NGC 4536 also has a starburst center \citep[e.g.][]{Davies1997}; 
it is possible that the overall higher temperature 
in starbursting galaxies could 
generate a strong enough gradient in temperature and \ttco \  optical
depth to produce an observable trend in \rtt \  compared to other galaxies.
The higher temperature might also account for the 
higher fluxes of \twco \ $(J = 2 \rightarrow 1)$ compared \twco \ $(J = 1 \rightarrow 0)$
spectra near the systemic velocity shown in Figure \ref{fig:spectra_sd}.

The stacked ratios in 5 galaxies, NGC 3593, 4254, 4273, 4654, and 5713, 
are larger than the weighted mean $\left< \mathcal{R}\right>$
beyond galactocentric distance of $\sim 1 \  \rm kpc$. 
The elevated \rtt \ in the outer regions, if due to decreased CO opacity, may reflect lower density in the gas, since the expected changes in temperature and line width (decreasing away from the galaxy center) would tend to {\it increase} the opacity. 
Alternatively, a positive radial gradient of $[\rm ^{12}C/^{13}C]$ 
abundance  could also result in higher \rtt \ in the outer region.

\subsubsection{Dependence on line width}

\begin{figure*}[ht]
\plotone{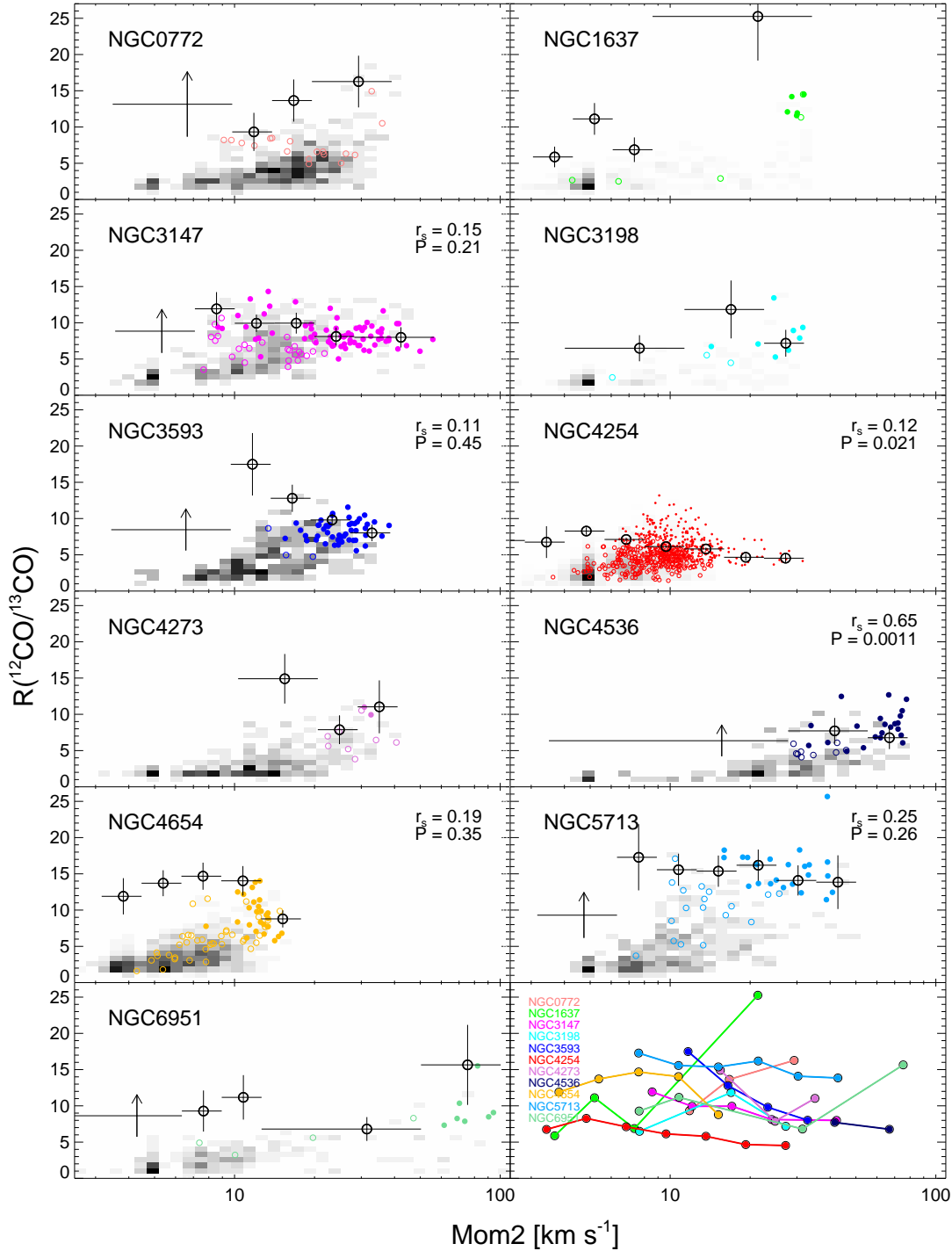}
\caption{
Line intensity ratio \rtt \ as a function of moment-2. 
The first 11 panels show the results for each individual galaxy, 
excluding NGC 1569.
Colored circles show the ratios for individual half-beams; 
the filled ones are those used for deriving 
Spearman's rank correlation coefficient $r_s$ and 
the significance $P$ shown in the top right of each panel. 
The gray scales show the distribution of 
\rtt$_{\rm min} = $\itw$/3\sigma_{13}$ for the \ttco\ non-detections. 
The black circles and arrows are the  stacked \rtt \ and lower limits of \rtt \ 
as a function of moment-2 respectively; 
the horizontal error bars reflect the bin size, and the vertical error bars show their uncertainties. 
The last panel is a summary plot showing trends of the stacked \rtt,
with different colors representing different galaxies.
}
\label{fig:rmom2}
\end{figure*}

Increasing the line width will tend to reduce the opacity in a single velocity channel. 
As long as the \twco \ line remains moderately optically thick, 
the larger velocity dispersion will elevate the \twco \ intensity and thus 
\rtt \ if the other properties of the molecular gas do not vary significantly. 
We used the M12 masks described in Section 3.1 to generate 
maps of the \twco \ intensity-weighted second moment (``moment-2"). 
The moment-2 value for each half-beam is a measure of 
the width of the spectral line ($\sim 0.42$ FWHM for a Gaussian profile).  
The range of the moment-2 for the samples is $\sim 5 - 100$
\kms, with the lowest line width determined by the spectral resolution 
of $10$ \kms \ we used. 
The line width in each galaxy generally decreases at larger 
distance from the galaxy center, except for NGC 772 where 
the moment-2 peaks in one of the spiral arms.

We measured the dependence of \rtt \ on moment-2 in
a similar manner as for galactocentric radius and show the results in Figure \ref{fig:rmom2}.
No strong correlations are found in any of the
galaxies for the resolved \rtt. 
In some galaxies, the stacked \rtt \ with lowest moment-2 show higher values. 
This trend is connected to the radial trend, with moment-2 decreasing with radius. 
For different galaxies, a larger moment-2 does not directly 
lead to a higher \rtt \ overall. 
These results imply that the changes in the \twco\ line width in galaxies 
do not lead to significant variations in \rtt.

\subsubsection{Effect of star formation rate}
\begin{figure*}[ht]
\plotone{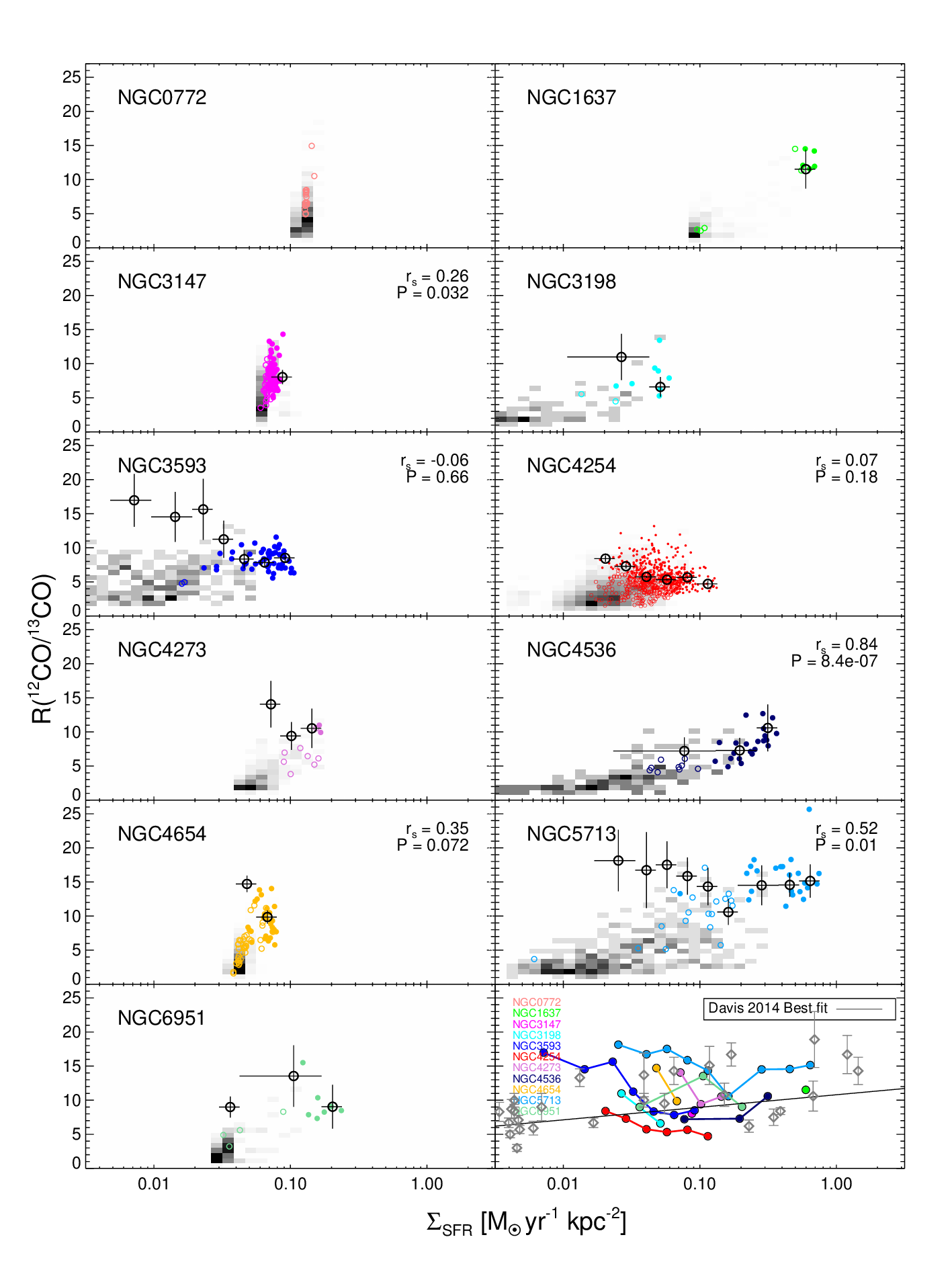}
\caption{
Line intensity ratio \rtt \ distribution as a function of SFR.
The first 11 panels show the results for each individual galaxy in 
the similar manner as Figure \ref{fig:rradius}. 
In the last panel, colored circles are the trends of stacked \rtt; 
the black solid line is the linear function provided by Equation
1 of \citet{Davis2014}, which results from fitting measurements of 
\rtt \ and star formation rate reported for galaxies in the literature; 
gray diamonds are the reported values for galaxies in Table 1 of 
\citet{Davis2014}.  
}
\label{fig:ri24}
\end{figure*}

Star formation activity in galaxies depends on the molecular gas 
properties, and also changes the physical conditions of the surrounding 
environment through feedback. 
\rtt, as a possible indicator of variations in molecular gas conditions, 
could vary with SFR. 
The large turbulence combined with the higher temperature 
and radiation strength in starbursts could explain their 
elevated \rtt \ values \citep{Aalto1995}. 
Gradients of \rtt \ decreasing away from local starbursting 
regions have also been seen in some  spatially 
resolved studies \citep{Aalto2010, Hirota2010, Watanabe2011}.

To study the correlation between \rtt \ 
and SFR density for individual half-beams, 
we use the $24 \rm \mu m $ intensity from Spitzer MIPS as a proxy for SFR.
For the 2 galaxies with common beam size larger than the $6 \arcsec$ resolution of MIPS $24 \rm \mu m $, NGC 3147 and 6951, we convolved their $24 \rm \mu m $ maps to match the common resolution of the \twco \ and \ttco.
The $24 \rm \mu m $ maps are 
further re-sampled onto the hexagonal grids described in Section \ref{sec:mom0}. 
We calculate the SFR following Equation 7 in \citet{Calzetti2010} as 
\beq
\mathrm{SFR}(M_{\odot} \ \mathrm{yr^{-1}}) = 2.46 \times 10^{-43} L_{24 
\mathrm{\mu m}} (\mathrm{ergs \ s^{-1}}).
\eeq

The results are shown in Figure \ref{fig:ri24}. 
Galaxies with large ranges of SFR density show that the stacked \rtt \ is higher for low SFRs. 
For the resolved \rtt,  only in NGC 4536 does \rtt \ increase with SFR; in all the other galaxies, \rtt \ does not show a strong correlation with the SFR. NGC 4536 is also the only galaxy showing a strong radial gradient of \rtt. Both of its correlations with SFR and radius can be explained by decreased opacity as a result of strong stellar feedback. We discuss the apparent lack of dependence of \rtt\ on SFR for other galaxies further in Section \ref{sec:davis}.

\section{Discussion}\label{sec:discussion}

\subsection{Line ratio and ISM density structure}\label{sec:diffuse}

There are both observational findings and theoretical predictions of 
the existence of diffuse gas only traced by \twco. 
In previous studies, \rtt \ values in the diffuse ISM are generally larger than 
those in bound, dense clouds (GMCs) \citep{Blitz1984, Knapp1988}.   
In the Milky Way, observations of diffuse gas find \rtt \ $\gtrsim$ 10--15, 
while \rtt \ as low as 3--4 has been reported for dark clouds 
\citep{Liszt2010}. 
In nearby galaxies, both GMCs and diffuse gas are included in 
the larger extragalactic beams, so 
\rtt \ of the beam depends on the ISM density structure. 
A higher fraction of diffuse gas in the beam 
will lead to a higher \rtt \ value than in the case of dense, 
opaque clouds \citep{Wilson1994}. 
On (sub-)kpc scales, the typical \rtt \ found in nearby galaxies is
$\sim$ 8 in \ttco \ surveys \citep[e.g.][]{Y&S1986,Paglione2001}, 
while \rtt \ $\gtrsim 20$ is often considered as an indicator of
diffuse gas \citep[e.g.][]{Aalto2010}.

In our sample, the spatially resolved \rtt \ of each galaxy  
are mostly in the range of 7--9, and no extremely high \rtt \ is found in our sample. 
However, our resolved \rtt \ measurements on kpc scales have two shortcomings:  
lack of short-spacing on large scales 
and limited sensitivity on small scales.  
First of all, 
our interferometer-only data may be insensitive to a spatially 
extended component of diffuse gas. 
Examining the effect of including single-dish data,
\citet{Pety2013} found diffuse molecular gas  
contributing $50\%$ of the \twco \ emission on scales larger than $1 \rm \  kpc$ in M51. 
Compared to \citet{Pety2013},  the galaxies in our sample are more distant than M51, 
and CARMA offers shorter spacings than the IRAM interferometer used in
their study. 
Based on the comparison of \twco$(J = 1 \rightarrow 0)$ and \twco$(J = 2
\rightarrow 1)$ flux presented in Section 3.2, 
we recover $50\% - 80\%$ of the flux in \twco \ with the interferometer data. 
Assuming all the missing flux is contained in diffuse gas 
that has intrinsically high \rtt \ values,  
the flux ratios of $F($\twco$)/F($\ttco$)$ could be underestimated by
up to a factor of 2 in our study.

On the other hand, even when diffuse emission is not resolved 
out by the interferometer, it may not be detected in individual half-beams
because of the limited brightness sensitivity of interferometer maps.
Because the interferometer maps usually have higher noise levels, 
low SNR emission will not be detected in individual half-beams. 
This can be inferred from the comparison between the 
weighted mean ratio $\left<\mathcal{R}\right>$ and the flux ratio  $F($\twco$)/F($\ttco$)$ 
in Figure \ref{fig:rvsfr}:
the former, resulting directly from the \ttco-detections, 
are generally less than the flux ratios, which include the low SNR regions.
The different relative ratios of $F($\twco$)/F($\ttco$)$ to  $\left<\mathcal{R}\right>$ 
in the galaxies can be interpreted as indicating different contributions 
of diffuse gas to $F($\twco$)/F($\ttco$)$. 
In NGC 772,  4273, and  4654, $F($\twco$)/F($\ttco$)$ is 
more than 3 times larger than $\left<\mathcal{R}\right>$, suggesting 
there is a significant amount of diffuse gas with higher line ratios distributed on large scales.
Furthermore, the stacking analysis allows us to measure 
average \rtt \ with better sensitivity and thus identify regimes in which diffuse gas may lead to higher \rtt.
Figure \ref{fig:i13vs12} shows
 that some galaxies with slightly enhanced $F($\twco$)/F($\ttco$)$ 
ratios show excess stacked \rtt \ at low \itw \ intensity, such as 
in NGC 3593, 4273, and 4654. 
This suggests that a diffuse component becomes dominant 
in low brightness regions. 
Moreover, we found that 8 galaxies in our sample (except NGC 1637, 4536 and 6951) show stacked \rtt \ at their largest galactocentric distances than \rtt \ of the innermost region we can probe. In 6 galaxies, NGC 772, 3198, 3593, 4254, 4273, 4654, and 5713, the stacked \rtt \ beyond a distance of $\gtrsim 1 \rm kpc$ are larger than the weighted mean $\left<\mathcal{R}\right>$.
In these 6 galaxies, only NGC 772 shows stacked \rtt \ increasing with velocity dispersion; 
for the other 5 galaxies, stacked \rtt \ does not depend on line width but decreases with SFR for $\Sigma_{\mathrm{SFR}}\lesssim 0.1 \un  M_{\odot} yr^{-1} kpc^{-2} $. 
It is possible that these trends are due to the increased fraction of diffuse gas at larger galactic radii; this is compatible with the expectation that diffuse gas becomes more abundant when going from center to atomic-dominated regions. 
Alternatively, it is also possible that the radial trend is drawn from a SFR trend; for NGC 3593, 4273, and 4654, it can also be a result of \rtt's dependence on \itw. 
We conclude that for normal spiral galaxies, half-beams with diffuse gas are most likely to be distributed in outer disk regions with low SFRs. 
However, as most of these half-beams are below our detection limit, 
higher sensitivity resolved \ttco \ mapping would be needed to 
better identify their actual distribution and correlations with other local properties.
Meanwhile,  
further studies incorporating single dish flux ratios 
are indispensable to identify extended diffuse structures.

\subsection{Lack of correlation between resolved \rtt \ and  SFR}\label{sec:davis}

Using global \rtt \ in more than 40 galaxies from literature, 
\citet{Davis2014} found \rtt \ is correlated with 
SFR surface density among spiral and early-type galaxies,  
and attributed this trend to
systematically 
higher gas temperature and/or velocity dispersion with increasing SFR. 
Higher star formation could heat the gas, and a higher 
temperature will reduce optical depths of both \ttco \ and \twco \  emission, 
resulting in a higher \rtt \ \citep{Aalto1995,Paglione2001}. 
Furthermore, feedback turbulence from active star formation  
could also reduce the optical depth and elevate the line ratio \citep{Aalto2010}. 

In our sample, except NGC 4536, we do not find a strong correlation between 
resolved \rtt \ and star formation rate within a galaxy 
based on a kpc-scale analysis (Figure \ref{fig:ri24}). 
One possibility is that the higher gas density associated with 
higher SFR surface density may offset the impact of temperature and
velocity dispersion on optical depth. 
Taking non-LTE effects into account, other studies using LVG models 
show that for molecular gas with high kinetic temperatures of $\sim 100 \rm \ K$, 
a volume density of $n_{\rm H_2} \sim 10^4 \rm \ cm^{-3}$ 
could result in \rtt \ $\sim 10$, comparable 
to \rtt \ values for typical conditions in the disk
\citep{Sakamoto1997, Meier2004}. 
For dense gas, the \ttco \ emission is enhanced compared to the 
lower density regions where it is subthermally excited, 
so \rtt \ measured in regions dominated by warm and dense gas could be
similar to that of the cold and lower density gas in typical disk conditions.   
The expectation of warm, dense gas in active star formation 
regions is also consistent with HCN observations
indicating large volume densities in such regions \citep[e.g.][]{Gao2004, Wu2005}. 
As discussed in Section \ref{sec:diffuse}, 
stacked \rtt \ decreases with SFRs 
for $\Sigma_{\mathrm{SFR}}\lesssim 0.1 \un  M_{\odot} yr^{-1} kpc^{-2}$ which might 
be due to the larger fraction of diffuse gas also 
also reflecting a density dependence for \rtt.
It is also interesting to note that in both of the two galaxies with 
relatively active in star formation 
in our sample, NGC 4536 and 5713, \rtt \ increases with SFR for 
$\Sigma_{\mathrm{SFR}}\gtrsim 0.1 \un  M_{\odot} yr^{-1} kpc^{-2} $.  
The offset of \rtt \ by density might be less for higher SFRs. 
However, more resolved observations are needed to confirm the trend.
On the other hand, if high temperature is not accompanied by higher 
gas density, as would be the case for diffuse gas adjacent to star forming 
regions, then higher \rtt \ would be expected; it is possible that 
such a trend may be missed by our interferometer data
which are less sensitive to this diffuse gas. 

Aside from modifying the optical depth, 
star formation activity can also affect \rtt \ by changing 
the abundance of \ttco \ relative to \twco.  
UV radiation from star formation sites can 
reduce the \ttco \ abundance by isotope-selective photodissociation, 
leading to higher \rtt \ in star forming regions.  
However, this photodissociation effect is generally disfavored by 
the lack of an observed correlation between \rtt \ and \ttco/$\rm C^{18}O$, 
given that $\rm C^{18}O$ should be more affected by isotope-selective photodissociation \citep{Paglione2001,Tan2011,Danielson2013}. 
Higher \rtt \ could also be due to an enhanced $\rm ^{12}C$ abundance 
resulting from recent massive star formation in star forming regions 
\citep{Casoli1991, Taniguchi1998}. 
We also did not find resolved \rtt \ increases with SFRs in each galaxy as expected 
if  \twco \ abundance is enhanced by isotope-selective photodissociation or massive star formation. 
In contrast, chemical fractionation towards \ttco \ at temperatures less than 35 K 
would decrease the relative abundance of \twco \ to \ttco, 
decreasing \rtt \ if the clouds remain cold \citep{Chu1983, Milam2005}. 
This effect could also compensate the  changes of opacity on \rtt. 
Given these competing effects of abundance and opacity, it is clear that more direct 
measurements of isotopic abundances
as a function of star formation activity are needed in addition to the line ratios \rtt.  

\subsection{Implications for $X_{\rm CO}$}
Since \twco \ is a widely used tracer for molecular mass, 
understanding how the CO-to-$\rm H_{2}$ conversion factor 
$X_{\rm CO}$ varies in different environments is crucial
for extragalactic studies. 
With less optical depth than \twco, \ttco \ 
would appear to be a better tracer of column density. 
In Milky Way studies, \ttco \ has been used as an independent measurement 
of molecular gas column density to calculate $X_{\rm CO}$ \citep[e.g.][]{Pineda2008, 
Roman-Duval2010}. 
However, these calculations are based on a series of assumptions 
that cannot easily be generalized to extragalactic environments:
\ttco \ emission traces the same density structure as \twco, 
the abundance of \ttco \ is constant, and most importantly, 
the gas is under LTE. 
When observed on large scales in external galaxies, 
\twco \ emission can also arise from the intra-cloud
gas in addition to the dense clouds which account for most of the \ttco \ emission. 
A fixed abundance of \ttco \ is unlikely to be appropriate for 
galaxies that span a large range in properties such as metallicity and star
formation rate. Similarly, the simplified assumption of LTE is also questionable 
when a wide range of local conditions occur within the telescope beam.
And even where LTE does approximately hold, an estimate of the 
kinetic temperature is needed to constrain the excitation analysis. 

Despite these difficulties inferring the $X_{\rm CO}$ factor from 
\ttco \ observations following LTE assumptions, the $X_{\rm CO}$ and \rtt \ are expected to be anti-correlated provided the
relative [\twco/\ttco] abundance does not change dramatically.
Both $X_{\rm CO}$ and \rtt \ depend on the bulk properties 
of molecular clouds; variations in physical conditions that affect one of them 
will also likely affect the other. 
For example, if an unusually high \rtt \ ($>20$) is identified such as 
in the case of a starbursting galaxy, 
there are good reasons to 
doubt if a standard $X_{\rm CO}$ should be applied. 
The $X_{\rm CO}$ values measured in starbursting galaxies are indeed $\sim 5$ times lower 
than the standard value \citep{Bryant1999, Papadopoulos2012}.
A larger velocity dispersion and higher temperature can explain the 
higher \rtt \ and lower $X_{\rm CO}$ at the same time. 
However, it is still difficult to observe a clear (anti-)correlation 
between \rtt \ and $X_{\rm CO}$  for normal galaxies. 
Among different galaxies, none of the properties 
we have examined (metallicity, inclination, line width and star formation activity)
seem to play an important role in determining the global \rtt. 
This is consistent with the nearly constant $X_{\rm CO}$ observed over 
a similar metallicity range in other studies. 
By inferring the $\rm H_{2}$ mass from the difference of dust-derived gas mass 
and HI observations, Sandstrom et al. (2013) reported a 
generally flat radial distribution for $X_{\rm CO}$ in nearby galaxies. 
We also find \rtt \ to be rather constant within a galaxy, 
which seems to be consistent with their results. 
While Sandstrom et al. (2013) found a lower $X_{\rm CO}$ in 
the central kpc in their sample galaxies on average,  
the lack of \rtt \ measurements outside $\sim 2 \rm \ kpc$ 
for most of our galaxies (due to sensitivity limitations)  
prevents us from comparing the outer region with the central kpc of our sample. 
In the future, use of additional diagnostics of the physical
conditions in the molecular gas (for example, high critical density
lines or higher excitation transitions) should permit a more detailed
description of variations in $X_{\rm CO}$.

\section{Conclusions}
We have measured \ttco \  fluxes for 12 galaxies in the CARMA STING sample and obtained \ttco \  maps for 11 of these on scales of a few hundred pc to a few kpc. 
We compare these maps with \twco \ observations, and investigate 
the resulting  \twco \ to \ttco \ line intensity ratio (\rtt) maps to study 
their dependence on line width, galactocentric distance and 
star formation activity. 
We also study the relation between galaxy properties and 
line ratios for galaxies in the sample.
Our main conclusions are: 

\begin{enumerate}
\item 
For the regions where both \twco \ and \ttco \ are detected, the resolved \rtt \ values range from $5.9$ to $14.9$, 
with most values between $7$ and $9$, 
which are quite similar to \rtt \ of GMCs in nearby galaxies.  
We do not find any half-beam with very high \rtt \ ($>20$) which are usually 
associated with presence of diffuse gas, 
implying that the majority of regions sampled in our data are 
likely to be dense gas structures.
We note that detection of very high resolved \rtt \  may require more sensitive observations.

\item For regions with high SNR of \ttco, the resolved \rtt \ in a galaxy increases with \itw. 
However, stacking all the beams with \twco \ detection regardless of \ttco \ SNR,  we found that the average \rtt \ does not depend on \itw, or show a decreasing trend. 
The discrepancy between the resolved \rtt \ and stacked \rtt \   can be explained as a bias
induced by the limited sensitivity in the resolved \rtt \ measurements.

\item Integrating \ttco \ over the entire galaxy, we obtain 
the flux ratios of \twco \ to \ttco \ emission ($F($\twco$)/F($\ttco$)$) 
for the 12 galaxies in our sample, with  a typical value around 10. 
By including half-beams under the \ttco \ 
detection threshold, the $F($\twco$)/F($\ttco$)$ is generally larger 
than the \ttco \ intensity weighted mean of the resolved line ratio $\left<\mathcal{R}\right>$ 
in a galaxy.  
Although there are significant galaxy-to-galaxy variations in both 
$F($\twco$)/F($\ttco$)$ and $\left<\mathcal{R}\right>$, 
we do not find clear dependences of the line ratios on galaxy properties.

\item 
Contrary to expectations from 
previous findings from single dish observations on galaxy scales, 
we find no clear evidence that resolved \rtt \ on (sub)kpc scales depends 
on the local properties of galactocentric radius, line width,  or star formation activity.
However, our resolved \rtt \ values are confined to regions with bright \ttco \ and hence span a small dynamic range in each of the properties. 
Nevertheless, our findings suggest that the effect of SF feedback on resolved \rtt \ on (sub)kpc scales 
may be compensated by the underlying density and/or an isotopic abundance gradient.

\item 
By stacking spectra for regions with low SNR of \ttco, we obtained the average \rtt \ as functions of galactocentric radius, line width, and SFR. NGC 3593, 4254, 4273, 4654 and 5713 show stacked \rtt \ that are elevated beyond  galactocentric distance of $\sim 1 \ \rm kpc$ as well as higher stacked \rtt \  with lower SFRs. We suggest that the increased \rtt \  are due to larger fractions of diffuse gas beyond galactocentric distance of $\sim 1 \ \rm kpc$.  Alternatively, stacked \rtt \ can also be increased  in outer regions if the \ttco \ abundance decreases away from the center. 
 
\item  
While resolved \rtt \ on (sub-)kpc scales 
show limited variations within a galaxy,   
indications of systematic differences of \rtt \ between galaxies 
are seen in our sample, implying that resolved \rtt \ 
might be sensitive to  
environmental properties, on local
or global scales, that are not captured by
our current analysis.  Inclusion of additional
data, such as stellar mass, star formation
and gas accretion history, or abundances of 
secondary elements, may be needed to better 
understand these variations.
 
\end{enumerate}

\acknowledgments
We thank the anonymous referee for helpful comments. 
We thank Katherine Alatalo and David Meier for helpful discussions. 
YC and TW acknowledge support from NSF through grants AST-1139950 and AST-1616199. 
ER is supported by a Discovery Grant from NSERC of Canada.
Support for CARMA construction was derived from the Gordon and Betty Moore Foundation, the Eileen and Kenneth Norris Foundation, the Caltech Associates, the states of California, Illinois, and Maryland, and the NSF. Funding for CARMA development and operations were supported by NSF and the CARMA partner universities. 
\bibliographystyle{apj}
\bibliography{sting13co}

\end{document}